\documentclass[%
 reprint,
nofootinbib,
 amsmath,amssymb,
 aps,
]{revtex4-2}

\usepackage{graphicx}% Include figure files
\usepackage{dcolumn}% Align table columns on decimal point
\usepackage{bm}% bold math
\usepackage{xr-hyper} % \hyperref[dm-...]{...}
\usepackage[colorlinks=true,allcolors=blue,bookmarks]{hyperref}

\usepackage{orcidlink}

\usepackage{ulem} % \sout

\newcommand{\HSout}[1]{\ifmmode\text{\color{red}\sout{\ensuremath{#1}}}\else{\color{red}\sout{#1}}\fi} % Henri's striked out comments

\newcommand{\pd}{\partial}
\newcommand{\rhoDM}{\rho_{\rm DM}}
\newcommand{\vDM}{v_{\rm DM}}
\newcommand{\Mso}{M_{\star}} % present-day stellar mass
\newcommand{\UniverseMachine}{UniverseMachine }%\textsc{UniverseMachine}

\newcommand{\appsection}[1]{\section{\MakeUppercase{#1}}}

\newcommand\aap{Astron.~Astrophys.}
\newcommand\apjl{Astrophys.~J.~Lett.}
\newcommand\apjs{Astrophys.~J.~Suppl.~Ser.}
\newcommand\mnras{Mon.~Not.~R.~Astron.~Sco.}
\newcommand\jcap{J.~Cosmo.~Astropart.~Phys.}
\newcommand\pasp{Publ. Astron. Soc. Pac.} % Publications of the Astronomical Society of the Pacific
\newcommand\physrep{Phys.~Rep.} % Physics Reports
\begin{document}

\preprint{APS/123-QED}

\title{Dark-matter-induced transients over cosmic time: The role of star formation history profiles}

\author{H. Steigerwald \orcidlink{0000-0001-6747-7389}}
%\email{heinrich.steigerwald@ufes.br}
\email{heinrich@steigerwald.name}
%\affiliation{Center for Astrophysics and Cosmology (Cosmo-ufes) and Department of Physics, Federal University of Esp\'{i}rito Santo, 29075910 Vit\'{o}ria, Esp\'{i}rito Santo, Brazil.}%Lines break 
\affiliation{PPGCosmo \& Cosmo-ufes, Universidade Federal do Esp\'{i}rito Santo, 29075-910, Vit\'{o}ria, ES, Brazil}

\date{May 27, 2025}

\begin{abstract}
The dark matter (DM) conundrum is one of the most intriguing due to its resistance in direct detection experiments. In recent years, attempts to identify non-gravitational signatures as the result of DM traversing or accumulating within stars have attracted a lot of attention. 
These calculations are usually evaluated at the order-of-magnitude level for stellar populations where the DM density is highest, such as galactic centers. 
However, if the signature implies the destruction of the host star, their population could have been diminished over a Hubble time in the most DM-dense regions, unless replenished by star formation. 
This circumstance exemplifies the need for galactic star formation history profiles when deriving DM-induced transient rates, in particular for predicting the host-offset distribution.
Here, we combine theoretical and empirical scaling relations of galaxy structure, star formation, and stellar initial mass function to construct a simple and efficient framework that permits us to estimate the target population formation rate and mass function within galactocentric radial zones across galaxy stellar masses and cosmic time. 
In a companion paper, we apply the framework to the hypothesis that DM in the form of primordial black holes accounts for the ignition of normal type Ia supernovae when colliding with white dwarf stars. 
\end{abstract}

\maketitle

%\tableofcontents

%%%%%%%%%%%%%%%%%%%%%%%%%%%%%%%%%%%%%%%%%%%%%%%%%%%%%%%%%
%%%%%%%%%%%%%%%%%%%%%%%%%%%%%%%%%%%%%%%%%%%%%%%%%%%%%%%%%
\section{Introduction}\label{sec:introduction}
The quest to unravel the nature of dark matter (DM) is shared between direct detection experiments and non-gravitational astrophysical signatures in stellar objects. 
Such signatures are particularly interesting in compact stars, where different mechanisms can lead to the complete destruction of the star producing powerful transient events detectable over cosmological distances. 
As for white dwarfs (WDs), explosion as a thermonuclear supernova (SN Ia) may be caused either by the encounter with an asteroid-mass primordial black hole (PBH)
%\cite{Graham:2015apa,
\cite{2025arXiv250521256S,2021PhRvL.127a1101S,2015PhRvD..92f3007G,*2019JCAP...08..031M,*2022MNRAS.510.4779S,*2024PhRvL.132o1401S} or as the result of continuous accumulation of heavy asymmetric particle DM %\cite{Bramante:2015cu}
\cite{2015PhRvL.115n1301B,*2018PhRvD..98k5027G,*2019PhRvD.100d3020A,*2019PhRvD.100c5008J}; in the latter case, implosion of the star to a solar-mass black hole is an alternative, producing a short electromagnetic burst, and possibly leading to solar-mass Ligo-Virgo-Kagra events in the case of binary systems \cite{2022PhRvD.105h3507S}. 
As for neutron stars (NSs), implosion to a black hole can be caused either by the capture of a PBH \cite{2013PhRvD..87l3524C,*2014JCAP...06..026P,*2020PhRvD.102h3004G}, or by continuous accumulation of asymmetric DM \cite{2012PhRvD..85b3519M}, both situations leading to the emission of non-repeating fast radio bursts and Ligo-Virgo-Kagra events in the case of binary systems. Different forms of composite DM structures \cite{2019PhRvD..99h3008G,*2021PhRvD.103l3022A,*2024PhRvD.109l3020R} have also been proposed to produce stellar destructions and transient events (see Ref.~\cite{2024PhR..1052....1B} for a comprehensive review of DM effects on compact stars).

DM may also settle in bound orbits upon star formation and lead to the implosion of the host star once that enters the remnant (WD or NS) phase
% Capela, Pshirkov, Tinyakov (2013)
\cite{2013PhRvD..87b3507C,*2014PhRvD..90h3507C}. 
In dwarf galaxies, the capture rate upon star formation is particularly high and may already destroy a significant fraction of stars during their main-sequence life time \cite{2022MNRAS.517...28O,2023PhRvD.107j3052E,*2024MNRAS.529...32E}. 
In case of main-sequence stars, the signatures may include a permanently visible impact on the present-day stellar mass function in the most extreme scenarios \cite{2023PhRvD.107j3052E,*2024MNRAS.529...32E}. 
However, in the case of stellar remnants, which are dim and whose presence can't be proven directly beyond a few hundred parsecs %Kilic
(e.g. Ref.~\cite{2020ApJ...898...84K}), the main observational signature is the transient. 

Since the rate of capture or encounter is proportional to  DM density and inversely proportional to DM velocity dispersion, the transient %implosion/explosion
rate per compact star is proportional to $\rhoDM/\vDM$. Therefore, regions where $\rhoDM/\vDM$ is highest---such as galactic centers---set the tightest constraints on DM parameter spaces, when compared with the allowed maximum rate derived from observations. 
However, this is not necessarily the case, if the rate is larger than the inverse of the Hubble time, whereupon the compact star population suffers significant depletion. Coincidentally, in view of inside-out formation of galaxies, regions of strongest depletion---i.e. galactic centers---are least replenished.

Consider, as an example, DM ignitions of SNe Ia. Owing to their high rate, progenitor WD depletion applies. Consequently, the recurrent argument that SNe Ia can't be ignited by DM because they trace the stellar population density, is mistaken. Following the argument, otherwise the host-offset distribution of SNe Ia should be skewed towards galactic centers in the proportion of the DM density profiles. The pitfall is that, over cosmic history, the progenitor WD population becomes imbalanced with respect to the stellar population density, roughly in the inverse proportion of the DM density profile. 
Curiously, in dwarf galaxies, the effect is regulated by DM-induced baryonic feedback \cite{2023arXiv230908661A}, leading to a complex interplay between gas infall, star formation, explosions, gas outflow, and gravitational-potential variations.

These considerations call for a more careful treatment of star formation histories (SFHs) in galactocentric radial zones when deriving restrictions on DM parameter spaces 
from DM-induced signatures that involve the destruction of stars.
Motivated by the hypothesis that PBH-WD encounters could be the origin of normal SNe Ia \cite{2025arXiv250521256S,2021PhRvL.127a1101S}, we attempt to fill this gap, in order to test weather or not the predicted rate distributions---in particular the host-offset---are compatible with the observed \cite{2025arXiv250521256S}. The task is helped by the fact that DM halos follow approximately spherical symmetry, which allows to model the stellar component---typically with disk-like symmetry---as if it was disposed in a spherically symmetric configuration too, since all that matters for our purpose is the DM-stellar encounter rate.
Using empirical galaxy size-mass-redshift scaling relations %(Van der Wel et al. 2014)
\cite{2014ApJ...788...28Vmax10}
and literature SFH models
% Behroozi  %Zahid et al. (2012), Childress et al. 2014, Sec.~2)
\cite{2013ApJ...770...57B,*2013ApJ...762L..31B,2011ApJ...734...48L,2012ApJ...757...54Z,2014MNRAS.445.1898C,2021MNRAS.506.3330Wmax10}, we developed a simple and efficient framework to 
%model
predict the SFH in galactocentric radial shells.

The paper is organized as follows: In \S~\ref{sec:galaxy-structure}, we review static galaxy structure relations and how to construct three-dimensional model galaxies from empirical sky-projected scaling relations. 
In \S~\ref{sec:GMA}, we review and perform galaxy mass assembly simulations. 
In \S~\ref{sec:radial-shell-modelling}, we combine the results of the first two sections to obtain a  radial-shell SFH model, where $[\pd\Psi(t;r,\Mso)/\pd r]\;\!dr$ denotes the SFH in a three-dimensional radial shell between galactocentric radii $r\to r\!+\!dr$ (hence integrated over angles) in an average galaxy with present-day stellar mass $\Mso$, related to the galaxy's total SFH by 
\begin{align}
    \Psi(t;\Mso) = \int\!\frac{\pd\Psi}{\pd r}(t;r,\Mso)\;\!dr\,.
\end{align}
We discuss our results in \S~\ref{sec:discussion}. Finally, we supply an \hyperref[sec:WD-formation]{Appendix}
where we compute the remnant mass function (focusing on SN Ia progenitors) for any instant after star formation and under various assumptions of the stellar initial mass functions (IMF) 
and initial-final mass relation. 

Throughout the paper, we use fixed cosmology with $\Omega_m=0.3$, $\Omega_{\Lambda}=0.7$ and $H_0=70~$km~s$^{-1}$Mpc$^{-1}$. Redshift and cosmic time are used interchangeably, which, for a matter-plus-cosmological constant universe, are linked by the mapping
\begin{align}
    t = \frac{2}{3\sqrt{\Omega_{\Lambda}}\;\! H_0}\ln\Big[\Big(\frac{a}{a_{m\Lambda}}\Big)^{3/2} + \sqrt{1+\Big(\frac{a}{a_{m\Lambda}}\Big)^3} \Big]\,,
\end{align}
where $a=(1+z)^{-1}$ is the scale factor, and $a_{m\Lambda} = (\Omega_{m}/\Omega_{\Lambda})^{1/3}$ is the scale factor at matter-cosmological constant equality. We note $\log(x)\equiv \log_{10}(x)$ and $\ln(x)\equiv \log_e(x)$.

%===========================================================
\section{Static galaxy structure scaling relations}\label{sec:galaxy-structure}

\begin{figure}
    \centering
    \includegraphics[width=0.9\linewidth]{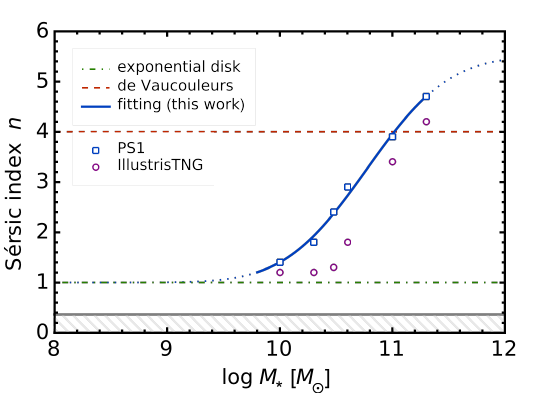}
    \caption{Median trend of the Sérsic index with stellar mass according to observational data at $z\simeq 0.05$ from PS1 survey %Pan-STARRS1 (Chambers 2016) % arxiv ok
    \protect\cite{2016arXiv161205560Cmax10} and numerical simulation results from IllustrisTNG % (Rodriguez-Gomez et al. 2019, Fig.~7)
    (Fig.~7 of Ref.~\protect\cite{2019MNRAS.483.4140R}). Fitting to PS1 data (full blue) with formula~\eqref{eq:nSersic-fit} and extrapolation (dotted blue) is also compared to exponential disc (green dot-dashed) and de Vaucouleurs % (1948)
    \protect\cite{1948AnAp...11..247D} profile (red dashed). The hatched region indicates where the $b_n$-approximation of eq.~\eqref{eq:bn-Sersic} is not valid.}
    \label{fig:nSersic}
\end{figure}

%----------------------------------------------------------
\subsubsection{Projected density profile}\label{sec:projected-density-profile}
The projected light intensity (or mass density) profile of individual galaxies is usually well-reproduced by the sum of a disc component and a bulge component, each parametrized by a Sérsic % (1963, 1968)
\cite{1963BAAA....6...41S,*1968adga.book.....S} profile
\begin{align}\label{eq:sersic}
    \Sigma(R) = \Sigma_0 \exp\Big[\!-\!b_n\Big(\frac{R}{R_{1/2}}\Big)^{\!\!1/n}\Big]\,,
\end{align}
where $R$ is the projected galactocentric radius coordinate perpendicular to the line of sight, $n$ is the Sérsic index (indicating the cuspiness), $R_{1/2}$ is the radius containing half of the projected luminosity\footnote{Since in mid-infrared wavelengths (such as \textit{Spitzer} \cite{2010PASP..122.1397Smax10} $3.6~\mu$m), half of the projected luminosity  corresponds to half of the projected stellar mass, we use the terms half-light radius and half-mass radius interchangeably.}, $\Sigma_0$ is the central surface density, and $b_n$ is the solution of $\Gamma(2n) = 2\;\!\gamma(2n,b_n)$, where $\gamma(a,x)\equiv \int_0^xt^{a-1}e^{-t}dt$ is the usual lower incomplete gamma function. A useful approximation for $b_n$, valid for $n>0.36$, is
\begin{align}\label{eq:bn-Sersic}
    b_n \simeq 2n-\frac{1}{3}+\frac{4}{405n}\,.
\end{align}

The profile is usually fitted separately for disc and bulge components, but it is also common---and more useful for our purpose---to fit the total galaxy (disk+bulge) profile by a single Sérsic profile, at the cost of precision loss for individual galaxies, but with gain of insight into overall scaling relations. According to observations from Pan-STARRS (PS1) \cite{2016arXiv161205560Cmax10} and numerical simulations \cite{2019MNRAS.483.4140R}, the median Sérsic index transitions roughly from $n\simeq 1$ at the low-mass end to $n\simeq 5.5$ at the high-mass end, with a relatively sharp transition occurring between $\Mso \simeq 10^{10}M_{\odot}$ and $\Mso\simeq 10^{11}M_{\odot}$ (see Fig.~\ref{fig:nSersic}), and has no noticeable dependence on redshift. We fit the median PS1 %Pan-STARRS
data %(Rodriguez-Gomez et al. 2019)
\cite{2019MNRAS.483.4140R} with an error function,
\begin{align}\label{eq:nSersic-fit}
    n = 3.25+2.25\;{\rm erf}\Big[\frac{\log(\Mso/M_n)}{\sigma_n}\Big]\,,
\end{align}
with transition mass $M_n$ and sharpness $\sigma_n$ as free parameters. Our fitting results are $\log(M_n/M_{\odot}) = 10.77$ and $\sigma_n = 0.81$ (see Fig.~\ref{fig:nSersic}).

\begin{figure}
    \centering
    \includegraphics[width=0.9\linewidth]{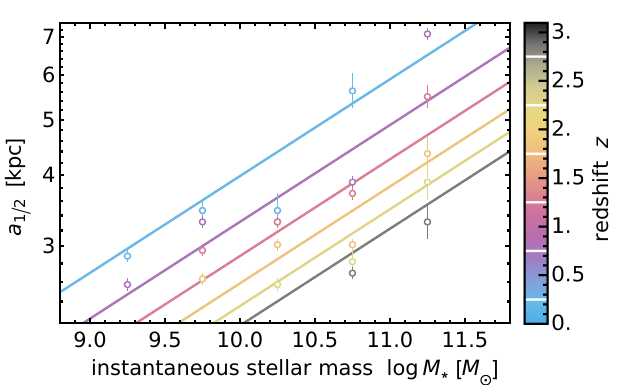}
    \caption{Median trend of the semi-major projected half-light radius with stellar mass according to CANDELS data %Van der Wel et al. 2014)
    \protect\cite{2014ApJ...788...28Vmax10} (colored circles) and fitted with scaling relation~\eqref{eq:half-light-radius} (colored lines).}
    \label{fig:ReSersic}
\end{figure}

The usual technique to measure the projected size of a galaxy is to suppose an (apparent) elliptical shape and determine major and minor half-light radii $a_{1/2}$ and $b_{1/2}$. In this case, the profile of eq.~\eqref{eq:sersic} is generalized from simple dependence on projected radius $R$ to dependence on projected major-axis and minor-axis radii $a$ and $b$, respectively. The usual empirical parametrization for the projected major-axis half-light radius is
\begin{align}\label{eq:half-light-radius}
    a_{1/2} = a_{1/2,0} \;\!\Big(\frac{\Mso}{10^{10}M_{\odot}}\Big)^{\!\alpha}(1\!+\!z)^{\beta}\,,
\end{align}
where $a_{1/2,0}$, $\alpha$ and $\beta$ are fitting constants and $\Mso$ is the \textit{instantaneous} stellar mass\footnote{\textit{instantaneous} stellar mass refers to the observed stellar mass of a galaxy, as opposed to its \textit{present-day} stellar mass which refers to the stellar mass it would have today} of the galaxy at redshift $z$. 
This relation is usually fitted separately for early- and late-type galaxies, which has been done by %Van der Wel et al. 2014)
Ref.~\cite{2014ApJ...788...28Vmax10} using CANDELS data covering the range $0<z<3$ and $9<\log(\Mso/M_{\odot})<12$. 
Since our concern is to  model volumetric DM-stellar encounter rates, we use the combined (early- and late-type) CANDELS data (table~5 of Ref.~%Van der Wel et al. 2014)
\cite{2014ApJ...788...28Vmax10}) and obtain the following fitting parameters for eq.~\eqref{eq:half-light-radius}: $a_{1/2,0}=4.506$~kpc, $\alpha = 0.17$, $\beta = -0.55$ (see Fig.~\ref{fig:ReSersic}).

In terms of the major-axis and minor-axis half-mass radii, the circularized projected half-mass radius is defined
\begin{align}\label{eq:circularized-half-light-radius}
    R_{1/2} = \sqrt{a_{1/2}\;\! b_{1/2}} \simeq 0.77~a_{1/2}\,, 
\end{align}
where in the second equality, we have used the observation that the average projected axis ratio is relatively independent of mass and redshift, $\langle b_{1/2}/a_{1/2}\rangle \simeq 0.6$ % (Van der Wel et al. 2014; Rodriguez-Gomez et al. 2019)
\cite{2014ApJ...788...28Vmax10,2019MNRAS.483.4140R}.

In the empirical approach of eqs.~\eqref{eq:half-light-radius} and \eqref{eq:circularized-half-light-radius}, the half-mass radius is a function of \textit{instantaneous} stellar mass and redshift, $R_{1/2}=R_{1/2}(\Mso,z)$, which is relatively inconvenient for describing the evolution of individual galaxies. Therefore, once the stellar mass histories $\Mso(t;M_{\star 0})$ are determined for a series of \textit{present-day} galaxy stellar masses $M_{\star 0}$ (in \S~\ref{sec:GMA}), we will rather make use of the half-mass radius history for a given \textit{present-day} stellar mass \footnote{we note that, whenever it is clear from the context, we will drop the subscript ``0" as indication for present-day}
\begin{align}\label{eq:circularized-half-light-radius-history}
    R_{1/2}(t;M_{\star 0}) \equiv R_{1/2}[\Mso(t;M_{\star 0}),z(t)]\,.
\end{align}

%--------------------------------------------------------------
\subsubsection{Three-dimensional density profile}
Deprojecting eq.~\eqref{eq:sersic} analytically for arbitrary Sérsic indices is mathematically rather challenging,
\begin{align}\label{eq:rho-star-int}
    \rho_{\star}(r) = -\frac{1}{\pi}\int_r^{\infty} \frac{d\Sigma}{dR}\frac{dR}{\sqrt{R^2-r^2}}\,.
\end{align}
Fortunately, a number of useful closed-form approximations for the integral~\eqref{eq:rho-star-int} have been proposed (see e.g. Ref.~\cite{2020A&A...635A..20V} % Vitral \& Mamon, 2020,
for a recent comparison). Here we chose the simple parametric form of %Pruguiel \& Simien%(1997)
Ref.~\cite{1997A&A...321..111P},
\begin{align}\label{eq:rho-star-Prugniel-Simien}
    \rho_{\star}(r) \simeq \rho_0\;\! \Big(\frac{r}{R_{1/2}}\Big)^{\!\!-p_n}\! \exp\Big[\!-\!b_n\Big(\frac{r}{R_{1/2}}\Big)^{\!\!1/n}\Big]\,,
\end{align}
where $p_n \simeq 1-0.6097/n+0.05463/n^2$ %(Lima Neto et al. 1999)
\cite{1999MNRAS.309..481L}, and $R_{1/2}$ is the projected half-mass radius (as determined previously in \S~\ref{sec:projected-density-profile}), and $\rho_0$ is the scale density.
According to % (Vitral \& Mamon 2020)
Ref.~\cite{2020A&A...635A..20V}, formula~\eqref{eq:rho-star-Prugniel-Simien} is accurate to within a few per cent with respect to the numerical integral of eq.~\eqref{eq:rho-star-int} for Sérsic indices $n\in[1,\;\!5.5]$ and projected radii $R/R_{1/2} \in [0.1,\;\!100]$, which is sufficient for our purpose. 

The (instantaneous) total stellar mass of a galaxy with profile~\eqref{eq:rho-star-Prugniel-Simien} is 
\begin{align}
    \Mso = 4\pi\!\! \int_0^{\infty}\!\!\!\rho_{\star}(r)\;\!r^2 dr = 4\pi \rho_0 \;\!R_{1/2}^{\,3}\;\!\frac{\Gamma[(3\!-\!p_n)\;\!n]}{b_n^{(3-p_n)n}}\,,
\end{align}
where $\Gamma(x)=\int_0^{\infty}t^xe^{-t}dt$ is the usual gamma function. This relation fixes the scale density in eq.~\eqref{eq:rho-star-Prugniel-Simien},
\begin{align}
    \rho_0 = \frac{\Mso\;\!b_n^{(3-p_n)n}}{4\pi\;\!R_{1/2}^{\,3}\;\!n\;\!\Gamma[(3\!-\!p_n)n]}\,,
\end{align}
and the (instantaneous) integrated mass up to an arbitrary radius $r$ has the form
\begin{align}\label{eq:Mstar-r}
    \Mso(r) = \Mso \frac{\gamma[(3\!-\!p_n)n,b_n(r/R_{1/2})^{1/n}]}{\Gamma[(3\!-\!p_n)n]}\,,
\end{align}
where $\gamma(x,a)=\int_0^xt^{a-1}e^{-t}dt$ is, again, the lower incomplete gamma function. Using this result, the (instantaneous) integrated mass between arbitrary radii $r_1$ and $r_2$ has the simple analytic form
\begin{align}\label{eq:Mstar-r1-r2}
    \Mso(r_1,r_2) \equiv &\; 4\pi\!\int_{r_1}^{r_2}\! r^2\rho(r)\;\!dr \nonumber \\
    =&\; \frac{\Mso}{\Gamma[(3\!-\!p_n)n]}\Big\{\gamma[(3\!-\!p_n)n,b_n(r_1/R_{1/2})^{1/n}] \nonumber \\ &\; -\gamma[(3\!-\!p_n)n,b_n(r_2/R_{1/2})^{1/n}] \Big\}\,,
\end{align}
where $\Mso$ is the total instantaneous stellar mass. Later on, in \S~\ref{sec:radial-shell-modelling}, we will make use of the integrated mass history, $\Mso(t;r_1,r_2,\Mso)$, between radii $r_1$ and $r_2$ for a given \textit{present-day} stellar mass $\Mso$, which is obtained by simply replacing in the right-hand side of eq.~\eqref{eq:Mstar-r1-r2},  $R_{1/2}\to R_{1/2}(t;\Mso)$, the half-mass radius history (see eq.~\ref{eq:circularized-half-light-radius-history}), and $\Mso\to \Mso(t;\Mso)$, the stellar mass history (which is the subject of the  next section).

%===========================================================
\section{`Point-like' galaxy mass assembly}\label{sec:GMA}

%-----------------------------------------------------------
\subsubsection{Mass-assembly framework}
We model the mean SFHs of galaxies following a procedure elaborated by %Leitner \& Kravtsov%(2011)
Ref.~\cite{2011ApJ...734...48L}, see also %Zahid et al. (2012), Childress et al. 2014, Sec.~2)
\cite{2012ApJ...757...54Z,2014MNRAS.445.1898C,2021MNRAS.506.3330Wmax10}, and the quenching prescription of %Childress et al.%Childress et al. 2014)
Ref.~\cite{2014MNRAS.445.1898C}, that we update slightly in order to be consistent with the most recent determination of the galaxy stellar mass function \cite{2020ApJ...893..111L,2018MNRAS.480.3491W,2014ApJ...783...85T}.
In this approach, the galaxy stellar mass increase rate is given by the difference between star formation rate (SFR)---mainly from dust and negligibly from accretion of dwarf galaxies---and the mass loss rate due to supernova explosions and stellar mass loss %(e.g.~Leitner \& Kravtsov 2011, Childress et al. (2014)
\cite{2011ApJ...734...48L,2014MNRAS.445.1898C}
\begin{align}\label{eq:dMsdt}
    \frac{d\Mso}{dt} = &\; \Psi\big[\Mso(t),z(t)\big] -\int_{t_f}^{t}\frac{df_{\rm ml}}{d\tau}\Big|_{t\!-t'}\;\!\Psi\big[\Mso(t'),z(t')\big]\;\!dt'\,,
\end{align}
where $\Psi(\Mso,z)$ is the (empirical) mean SFR at \textit{instantaneous} stellar mass $\Mso$ and redshift $z$,\footnote{Not to be confused with the SFH of a galaxy with \textit{present-day} stellar mass $\Mso$, noted $\Psi(t;\Mso)$, see eq.~\eqref{eq:SFH-def}.} and $f_{\rm ml}(\tau)$ is the fraction of stellar mass lost as a function of time $\tau$ after a star formation burst at $\tau = 0$, which, for the canonical %(Chabrier 2003)
\cite{2003ApJ...586L.133C}
stellar IMF, is %(Jungwiert et al., 2001; Leitner \& Kravtsov 2011)
\cite{2001A&A...376...85J,2011ApJ...734...48L}
\begin{align}\label{eq:mass-loss-rate}
    f_{\rm ml}(\tau) = 0.046\;\! \ln\Big(1+\frac{\tau}{0.276~{\rm Myr}}\Big)\,.
\end{align}
The star-forming main sequence's SFR is usually fitted by the parametric form
 (see, e.g., %Zahid et al. (2012)
 Ref.~\cite{2012ApJ...757...54Z}, even though other parametrizations are also in use, e.g., %Childress et al. 2014)
Ref.~\cite{2014MNRAS.445.1898C})
\begin{align}\label{eq:SMz-sf}
    \Psi_{\rm sf}(\Mso,z) = \;\!\Psi_{\rm sf,0}\;\!\Big(\frac{\Mso}{10^{10}M_{\odot}}\Big)^{\!\alpha} \;\!(1+z)^{\beta}, 
\end{align}
where $\Psi_{\rm sf,0}$, $\alpha$, and $\beta$ are constants, and $\Mso$ is the instantaneous stellar mass.

\begin{figure}
    \centering
    \includegraphics[width=\linewidth]{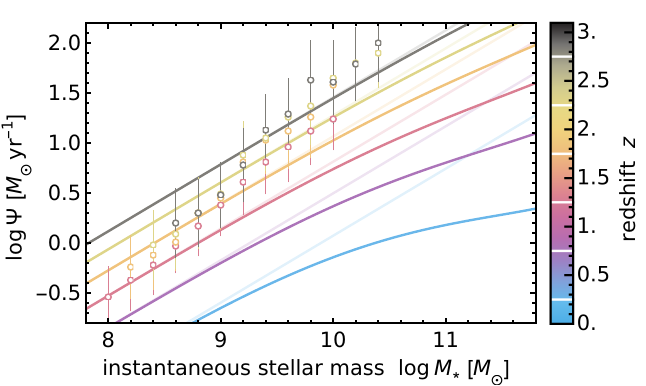}
    \caption{Median SFR assumed in this work with/without quenching (opaque/transparent lines) as a function of instantaneous stellar mass and for various redshifts as indicated by the bar chart. For comparison, circles show recent determinations of the star-forming main sequence from HST/GOODS-CANDELS/GTC/SHARDS surveys% (Mérida et al. 2023)
    \protect\cite{2023ApJ...950..125M}.}
    \label{fig:SFR}
\end{figure}

The average SFR of the total galaxy sequence, comprising star-forming and quiescent galaxies, can be written as %(Childress et al. 2014)
\cite{2014MNRAS.445.1898C}
\begin{align}\label{eq:SMz}
    \Psi(\Mso,z) = p_Q(\Mso,z)\;\!\Psi_{\rm sf}(\Mso,z)\,,
\end{align}
where $p_Q(\Mso,z)$ is called the `quenching penalty' function, which accounts for the gradual transition from active galaxies at low stellar mass to passive galaxies at high stellar mass, and can be parametrized as a simple error function %(Childress 2014)
\cite{2014MNRAS.445.1898C},
\begin{align}
    p_Q(\Mso,z) = \frac{1}{2}-\frac{1}{2}{\rm erf}\Big\{\frac{\log[\Mso/M_Q(z)]}{\sigma_Q}\Big\}\,,
\end{align}
with $\sigma_Q\simeq 1.5$ the transition width and the quenching mass scale determined observationally
(Fig.~A2 of Ref. % (Childress et al. 2014, eq.~A8)
\cite{2014MNRAS.445.1898C}),
\begin{align}\label{eq:quenching-mass}
    \log\big[M_Q(z)/M_{\odot}\big] = 10.077+0.636\;\!z\,.
\end{align}
The following alternative form of the quenching penalty function, introduced and advocated by Ref.~\cite{2014MNRAS.445.1898C}, has the advantage of allowing for a minimum value, %(Childress et al. 2014)
\begin{align}
    \tilde{p}_Q(\Mso,z) = 1-(z/10-1)^2[1-p_Q(\Mso,z)]\,.
\end{align}
The SFR including quenching, i.e.~eq.~\eqref{eq:SMz} with  $p_Q(\Mso,z)\to \tilde{p}_Q(\Mso,z)$, is shown as the opaque colored lines in Fig.~\ref{fig:SFR}, with values $\Psi_{\rm sf,0}=0.62M_{\odot}~$yr$^{-1}$, $\alpha=0.62$ and $\beta=2.9$, that we adopted pragmatically (fine-tuned) in order to reproduce the observed star formation rate density (SFRD, see \S~\ref{sec:SFRD}). Also shown in Fig.~\ref{fig:SFR} is the star-forming main sequence's SFR (transparent colored lines) compared to data from the HST/GOODS-CANDELS/GTC/SHARDS surveys % (Mérida et al. 2023)
\cite{2023ApJ...950..125M}. The match is rather crude but still within statistical error estimates.

\begin{figure}
    \centering
    \includegraphics[width=\linewidth]{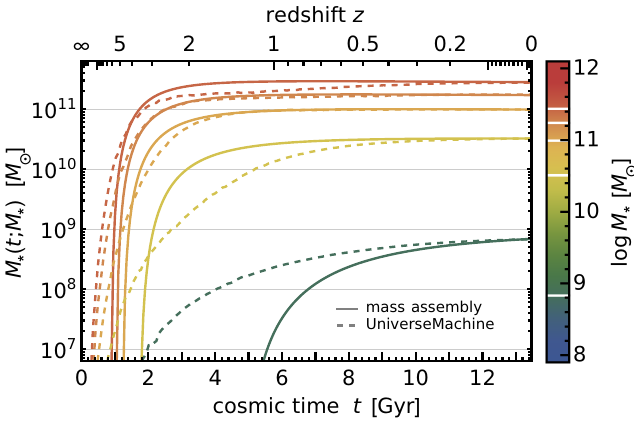}% produced by galaxy-mass-assembly.nb
    \caption{Stellar mass histories calculated with the mass assembly model (full lines) specified by eq.~\eqref{eq:dMsdt} and compared to \UniverseMachine simulation results (dashed lines) for a selection of present-day stellar masses as indicated by colors in the bar chart.}
    \label{fig:mass-history}
\end{figure}

With these ingredients, eq.~\eqref{eq:dMsdt} becomes a simple recurrent numerical scheme. For efficient numerical resolution, it is useful to transform the integral into a sum %Leitner \& Kravtsov%(2011)
\cite{2011ApJ...734...48L},
\begin{align}\label{eq:dMsdt-ML}
    M_{\star,n+1}=&\; M_{\star,n} + \Delta t\;\! \Psi(M_{\star,n},t_n)\nonumber \\
    &\; - \sum_{j=0}^n \Psi(M_{\star,n-j},t_{n-j})\;\!(f_{{\rm ml},j+1}-f_{{\rm ml},j})\,,
\end{align}
where $M_{\star,j}\equiv \Mso(t_f\!+\!j\;\!\Delta t)$, where $t_f$ is the formation time, $f_{{\rm ml},j}\equiv f_{\rm ml}(j\;\!\Delta t)$, and we use $\Delta t=0.5~$Myr (same as %Childress et al. 2014)
Ref.~\cite{2014MNRAS.445.1898C}). Following Ref. %Childress et al. 2014)
\cite{2014MNRAS.445.1898C}, we start the recurrent scheme with an initial stellar mass of $M_{\star,f}=10^6M_{\odot}$, and refer to this moment as the formation time, $t_f$. 
We perform stellar mass assembly simulations for present-day stellar masses in the range $\log(\Mso/M_{\odot}) \in [8,12]$ with logarithmic spacing of 0.2.
A selection of the reconstructed stellar mass histories $\Mso(t)$ can be visualized in Fig.~\ref{fig:mass-history}. 
The model sequences reproduce the well-established trend that lower mass galaxies form later, and that supermassive galaxies ($\log(\Mso)\gtrsim 11$) have already formed most of their stars by redshift $z \gtrsim 2$. 
Also shown in Fig.~\ref{fig:mass-history}, are average SFHs from \UniverseMachine  %(Behroozi et al. 2013a,b)
\cite{2013ApJ...770...57B,*2013ApJ...762L..31B}.
In comparison with these, low-mass galaxies form later and intermediate-mass galaxies form earlier. In addition, overall formation timescales are shorter, as will become clearer when comparing SFHs in the next section, Fig.~\ref{fig:SFH}.

%-----------------------------------------------------------
\subsubsection{Star formation histories}\label{sec:SFH}
\begin{figure}
    \centering
    %\includegraphics[width=\linewidth]{figs/SFHfitlog-1-1.pdf}
    % produced with galaxy-mass-assembly.nb
    \includegraphics[width=\linewidth]{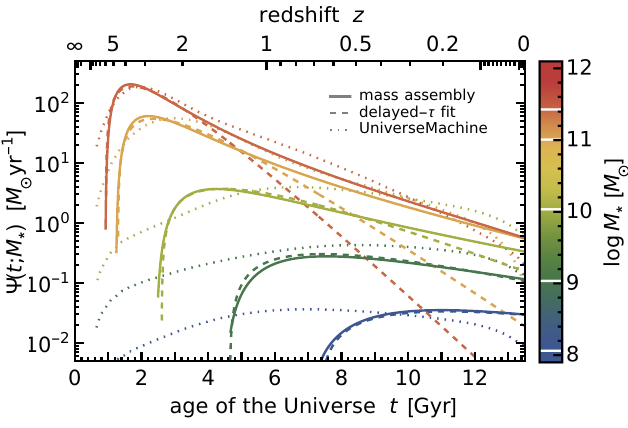}
    % produced with galaxy-mass-assembly.nb
    \caption{Star formation histories of model galaxies for a selection of present-day stellar masses (see color bar chart) as calculated by numerical integration scheme, eq.~\eqref{eq:dMsdt-ML} (full lines), and fitted with delayed-$\tau$ models, eq.~\eqref{eq:SFH-delayed-tau} (dashed lines), and also compared to \UniverseMachine%(Behroozi et al. 2013a,b)
    \protect\cite{2013ApJ...770...57B,*2013ApJ...762L..31B} average SFHs (dotted lines).}
    \label{fig:SFH}
\end{figure}

The SFH of a galaxy with present-day stellar mass $\Mso$ is mathematically the pullback of the (instantaneous) SFR by the stellar mass history, simply defined by (e.g. eq.~5 of %Leitner \& Kravtsov%(2011)
Ref.~\cite{2011ApJ...734...48L})
%\footnote{Note the typo in their eq.~5, where it should be SFH instead of SFR}
\begin{align}\label{eq:SFH-def}
    \Psi(t;\Mso) \equiv \Psi[\Mso(t;\Mso),z(t)]\,.
\end{align}
In Fig.~\ref{fig:SFH}, we show a selection of SFHs of the mass assembly model (full lines) and \UniverseMachine \cite{2013ApJ...770...57B,*2013ApJ...762L..31B}  (dotted lines).
The mass-assembly SFHs are comparable to those of % Childress et al. 2014, Sec.~2)
Ref.~\cite{2014MNRAS.445.1898C}, fig.~1.
%Zahid et al. (2012), Childress et al. 2014, Sec.~2)
%\cite{2012ApJ...757...54Z,2014MNRAS.445.1898C}
Some prominent similarities and differences of mass-assembly SFHs and \UniverseMachine SFHs can be noticed:
\begin{itemize}
    \item peak star formation epochs of the highest mass galaxies ($\Mso \gtrsim 10^{11}M_{\odot}$) are quite similar.
    \item mass-assembly  peak star formation epochs of intermediate mass galaxies ($10^{9}M_{\odot}\lesssim \Mso \lesssim 10^{10}M_{\odot}$) occur earlier and last shorter.
    \item mass-assembly peak star formation epochs of low-mass galaxies ($\Mso \lesssim 10^8M_{\odot}$) start later and are shorter.
\end{itemize}

\begin{figure}
    \centering
    \includegraphics[width=0.85\linewidth]{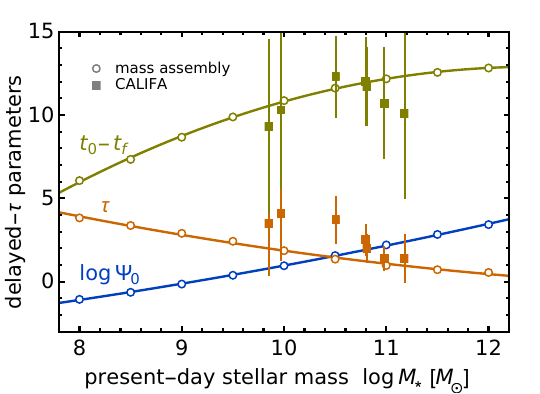}
    \caption{Delayed-$\tau$ fitting parameters in eq.~\eqref{eq:SFH-delayed-tau} for SFHs obtained with the mass assembly model of eq.~\eqref{eq:dMsdt} as a function of present-day stellar mass (colored circles), second-order fitting in $\log \Mso$ (colored lines), and compared to observations from the CALIFA survey (colored squares, taken from table~1 of Ref. %Lopez-Fernandez (2018) 
    \protect\cite{2018A&A...615A..27L}). The physical units are Gyr for $t_f$ and $\tau$, and $M_{\odot}$yr$^{-1}$ for $\Psi_0$, and $t_0=13.467$~Gyr is the present time.}
    \label{fig:delayed-tau-fit}
\end{figure}

For convenience, we fit the mass-assembly SFHs with delayed-$\tau$ models %(e.g., Bruzual \& Kron 1980; Chiosi 1980; Gavazzi et al. 2002; Lee et al. 2010; Simha et al. 2014, unpublished; Chiosi et al. 2017; López-Fernández et al. 2018)
\cite{1980ApJ...241...25B,*1980A&A....83..206C,*2002ApJ...576..135G,*2010ApJ...725.1644L,*2014arXiv1404.0402S,*2017ApJ...851...44C,2018A&A...615A..27L}, %SFR$(t)\propto (t_0-t)\exp[-(t_0-t)/\tau]$, 
\begin{align}\label{eq:SFH-delayed-tau}
    %\Psi_{\Mso}(t)
    %\Psi(t;\Mso) = \Psi_0 \frac{t_0-t}{\tau} \exp\Big[-\frac{t_0-t}{\tau}\Big]\,, \quad [M_{\odot}\,{\rm yr}^{-1}]
    \Psi(t;\Mso) \simeq \Psi_0\;\! \frac{t\!-\!t_f}{\tau}\;\! \exp\Big(\!-\frac{t\!-\!t_f}{\tau}\Big)\,,% \quad [M_{\odot}\,{\rm yr}^{-1}]
\end{align}
for $t>t_f$, and $\Psi=0$ otherwise, and where 
$\Psi_0$, $\tau$, and $t_f$ are fitting `constants' that depend on the present-day stellar mass $\Mso$ (see Fig.~\ref{fig:delayed-tau-fit}). 
Also shown in Fig.~\ref{fig:delayed-tau-fit} are observational determinations of the delayed-$\tau$ parameters from the CALIFA survey %Lopez-Fernandez (2018) 
\cite{2018A&A...615A..27L},
and we find good agreement.
Nevertheless, we note that the observed formation timescales $\tau$ (orange squares in Fig.~\ref{fig:delayed-tau-fit}) are slightly more extended for intermediate-mass galaxies ($\Mso \sim 10^{10}M_{\odot}$), consistent with \UniverseMachine (Fig.~\ref{fig:SFH}), suggesting that the mass-assembly SFHs probably slightly underestimate the real star formation extension at that mass scale (and probably at lower stellar masses too), even though the tension is minor.

\begin{figure*}[htb]
    \centering
    \includegraphics[width=0.6\linewidth]{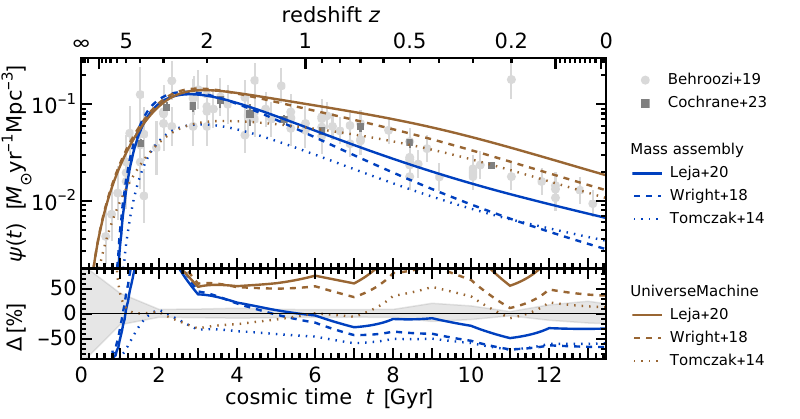}
    \caption{    
    Star formation rate density as defined by eq.~\eqref{eq:psi-GMA} for various galaxy stellar mass functions \protect\cite{2020ApJ...893..111L,2018MNRAS.480.3491W,2014ApJ...783...85T} (as indicated by line styles),
    for the mass assembly model SFR histories (blue lines) and \UniverseMachine%(Behroozi et al. 2013a,b)
    \protect\cite{2013ApJ...770...57B,*2013ApJ...762L..31B} SFHs (brown lines),
    compared to recent data from LOFAR %(Cochrane et al. 2023)
    \protect\cite{2023MNRAS.523.6082C} survey (dark gray squares) and a data compilation of Ref.~\protect\cite{2019MNRAS.488.3143B} (gray disks).
    The mini bottom panel shows relative errors (same line styles) with respect to %LOFAR%(Cochrane et al. 2023)\cite{2023MNRAS.523.6082C}
    data mean \protect\cite{2019MNRAS.488.3143B,2023MNRAS.523.6082C} and compared to its 1$\sigma$ confidence region (gray shaded area).}
    \label{fig:SFRD}
\end{figure*}

The delayed-$\tau$ fitted SFHs are also shown in Fig.~\ref{fig:SFH}. The fit is overall satisfying, but underestimates significantly the SFR of supermassive galaxies ($\Mso \gtrsim 10^{11}M_{\odot}$) at late times ($z \lesssim 1$); this should be acceptable for our purpose, because, as already mentioned when commenting Fig.~\ref{fig:mass-history}, these galaxies have already completed most of their star formation activity by redshift $z\gtrsim 2$.
For the average \UniverseMachine SFHs, we find that a double power-law provides a better fitting. The use of such fittings will become clearer in \S~\ref{sec:radial-shell-modelling}.

%-----------------------------------------------------------
\subsubsection{Star formation rate density}\label{sec:SFRD}
The main observational constraint on SFH models is the observed SFRD (see, e.g.,  %(Behroozi et al. 2013a,b)
Ref.~\cite{2013ApJ...770...57B,*2013ApJ...762L..31B}), 
defined as the integral over stellar masses pondered by the galaxy stellar mass function, $\phi(\Mso)$, 
\begin{align}\label{eq:psi-GMA}
    \psi(t) = \!\int\! \Psi(t;\Mso)\;\!\phi(\Mso)\;\!d\Mso\,,
\end{align}
where $\phi(\Mso)\;\!d\Mso$ represents the average number density of galaxies on comoving scales ($\gtrsim 100~$Mpc) with stellar mass in the range $\Mso\to \Mso\!+\!d\Mso$, 
and is usually fitted by a double %Schechter% (1976)
\citet{1976ApJ...203..297S} function % (Baldry et al. 2008, Pozzetti et al. 2010; Ilbert et al. 2013)
\cite{2008MNRAS.388..945B,*2010A&A...523A..13Pmax10,*2013A&A...556A..55Imax10}
\begin{align}\label{eq:GSMF}
    \phi(\Mso) =\frac{1}{M^*} \exp\Big(\!\!-\!\frac{\Mso}{M^*}\Big)\sum_{i=1}^2\phi_i^*\Big(\frac{\Mso}{M^*}\Big)^{\alpha_i}\!,
\end{align}
where $M^*$ is the characteristic Schechter mass, $\phi_1^*$ and $\phi^*_2$ are the characteristic comoving number densities of quiescent and star-forming galaxies, respectively, and $\alpha_1$ and $\alpha_2$ are their slopes, respectively.

In Fig.~\ref{fig:SFRD}, we compare the model SFRD, obtained for various determinations of the galaxy stellar mass functions \cite{2020ApJ...893..111L,2018MNRAS.480.3491W,2014ApJ...783...85T}, with recent data from the LOFAR survey %(Cochrane et al. 2023)
\cite{2023MNRAS.523.6082C} (dark gray squares) and a data compilation \cite{Behroozi:2019} %\verb|Behroozi:2019| 
%of Ref.\cite{2019MNRAS.488.3143B}
that was used to constrain \UniverseMachine simulations (gray disks).
\UniverseMachine SFHs (brown lines) best reproduce the SFRD with the galaxy stellar mass function of Ref.~\cite{2014ApJ...783...85T} (dotted brown line), while, by construction, the mass assembly SFHs (blue lines) best reproduce the observed SFRD with that of Ref. %(Leja et al., 2020)
\cite{2020ApJ...893..111L} (full blue line).
As can be seen in Fig.~\ref{fig:SFRD} bottom panel, the best-fit mass assembly SFRD (full blue line) is accurate to within a few tens of per cent overall, though it is a factor of $\sim 2$ higher at star formation noon ($z\sim 3$). The best-fit \UniverseMachine SFRD (dotted brown line) is overall accurate to within a few tens of per cent.\footnote{Let us quickly see how to evaluate the integral of eq.~\eqref{eq:psi-GMA} efficiently, when $\phi(\Mso)$ is given by eq.~\eqref{eq:GSMF}. First, observe that a finite integral has expression
\begin{align}\label{eq:GSMF-int}
    \int_{M_{\star,1}}^{M_{\star,2}} \phi(\Mso)\;\!d\Mso =&\; \sum_{i=1}^2\phi^*_i\Big[\Gamma\Big(\alpha_i\!+\!1,\frac{M_{\star,1}}{M^*}\Big) \nonumber\\
    \;-\Gamma\Big(\alpha_i\!+\!1,\frac{M_{\star,2}}{M^*}\Big)\Big]
\end{align}
where $\Gamma(a,x)=\int_x^{\infty}t^{a-1}e^{-t}dt$ is the usual upper incomplete gamma function.
Note that we can write eq.~\eqref{eq:psi-GMA} as a sum of $n$ `small' integrals
\begin{align}
    \psi(t) = &\; \;\! \sum_{j=1}^{n-1} \int_{M_{\star,j}-\Delta \Mso/2}^{M_{\star,j}+\Delta \Mso/2} \phi(\Mso)\;\! \Psi(t;\Mso)\;\!dM\,,
\end{align}
where the $M_{\star,j}$'s run from some lower limit $M_{\star,1}\sim 10^7M_{\odot}$ to some upper limit $M_{\star,n}\sim 10^{13}M_{\odot}$.
For sufficiently small intervals $\Delta \Mso$, and since $\Psi(t;\Mso)$ is a slowly varying function of $\Mso$ in any $[M_{\star,j}\!-\!\Delta \Mso/2, M_{\star,j}\!+\!\Delta \Mso/2]$, we can approximate $\Psi(t;\Mso)$ by its value at $M_{\star,j}$ and pull it out of the integral. Using eq.~\eqref{eq:GSMF-int}, we have 
\begin{align}
        \psi(t) \simeq &\;  \sum_{j=1}^{n-1}\Psi(t,M_{\star,j})\;\! \sum_{i=1}^2\phi^*_i\Big[\Gamma\Big(\!\alpha_i\!+\!1,\frac{M_{\star,j}\!-\!\Delta \Mso/2}{M^*}\Big) \nonumber \\
    &\;-\Gamma\Big(\!\alpha_i\!+\!1,\frac{M_{\star,j}\!+\!\Delta \Mso/2}{M^*}\Big)\Big]\,.
\end{align}
We find that accuracy of better than 1\% is achieved for $M_{\star,1}=10^7M_{\odot}$, $M_{\star,n}=10^{13}M_{\odot}$ and logarithmic intervals of at most $\Delta \log(\Mso/M_{\odot}) \lesssim 0.1$.}

\begin{figure*}[htb]
    \centering
    \includegraphics[width=0.32\linewidth]{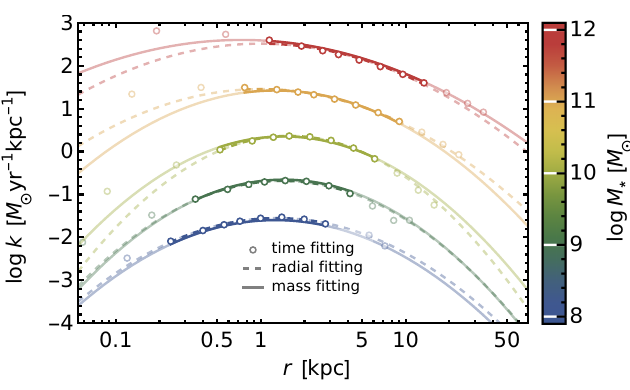}
    \includegraphics[width=0.32\linewidth]{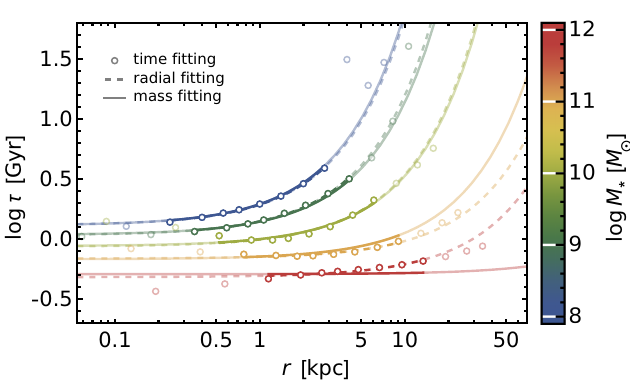}
    \includegraphics[width=0.32\linewidth]{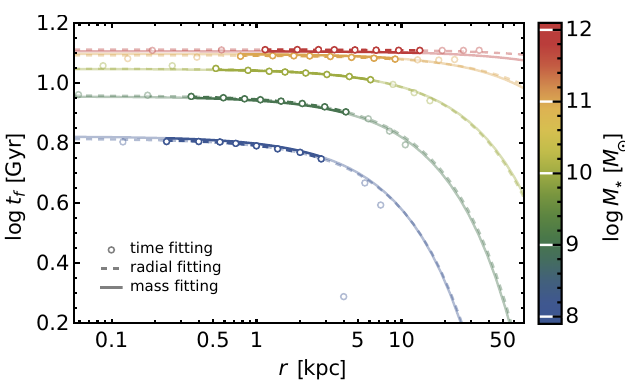}
    \includegraphics[width=0.32\linewidth]{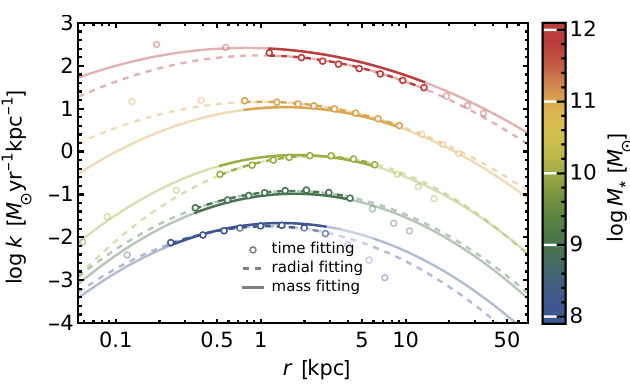}
    \includegraphics[width=0.32\linewidth]{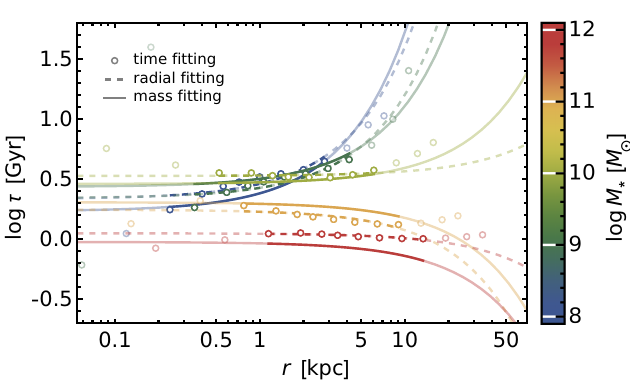}
    \includegraphics[width=0.32\linewidth]{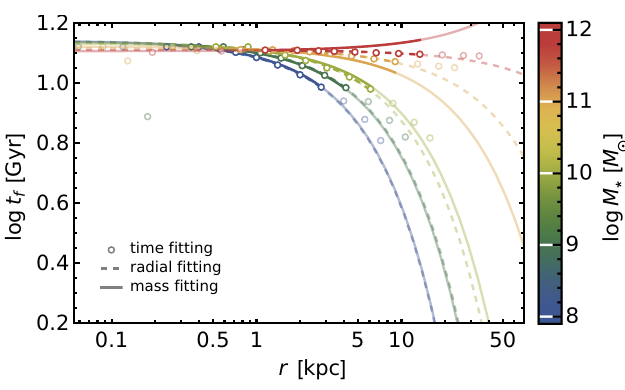}
    \caption{Radial fitting of delayed-$\tau$ parameters $k$, $\tau$ and $t_f$ of eq.~\eqref{eq:gold-formula} for a selection of present-day stellar masses as indicated by colors for SFHs from mass assembly (top) and \UniverseMachine (bottom). Transparent data points are excluded in order to optimize the result for the offset ranges where most stars form. Transparent lines indicate extrapolations. }
    \label{fig:fitrcomp}
\end{figure*}

\section{Galaxy mass assembly in radial zones}\label{sec:radial-shell-modelling}
With the preparing sections on galaxy structure (\S~\ref{sec:galaxy-structure}) and galaxy mass assembly (\S~\ref{sec:GMA}) in our baggage, we now have all the tools necessary to simulate SFHs within galactocentric radial shells. 
The procedure, new in the literature to our knowledge, is based on the assumption that eq.~\eqref{eq:dMsdt} holds when considering radial shells instead of entire galaxies.
This is justified because the net radial migration of stellar populations is close to zero for most parts of disc galaxies %(Frankel et al. 2018, Johnson+21)
\cite{2002MNRAS.336..785S,*2018ApJ...865...96F,2008ApJ...675L..65R,*2021MNRAS.508.4484J}.
In numerical simulations, however, very slow outward radial migration of stellar populations is observed in the outskirts for $r\gtrsim R_{1/2}$ \cite{2008ApJ...675L..65R,*2021MNRAS.508.4484J}. 
Here, we neglect these effects---including potential radial migration in elliptical galaxies which contribute about 10 percent to the SFRD---to focus on the main formation process of stellar populations in galaxies, and leave the treatment of secondary processes for future studies.
We also limit the framework, somehow arbitrarily, to within five present-day half-mass radii. 

\begin{figure*}[htb]
    \centering
    \includegraphics[width=0.45\linewidth]{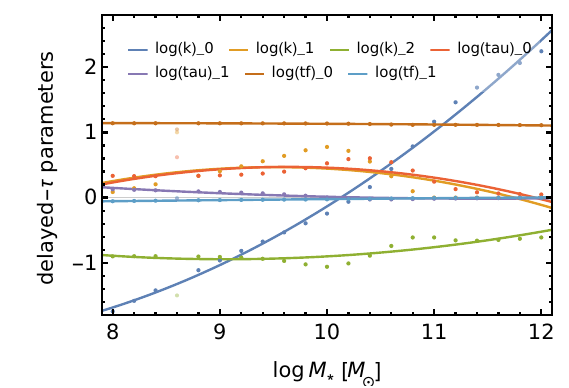}
    \includegraphics[width=0.45\linewidth]{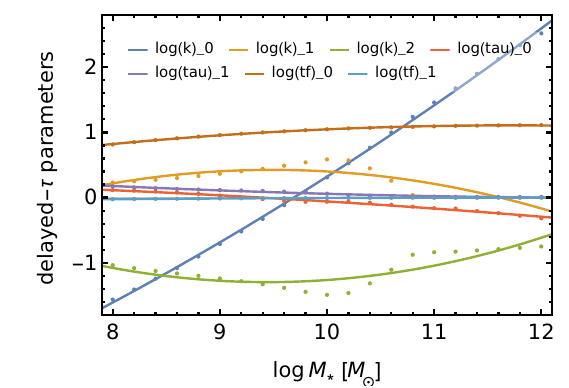}
    \caption{Fitting in $\log \Mso$ for SFHs from \UniverseMachine (left) and mass assembly (right). 
    Second-order fitting in $\log \Mso$ is sufficient to capture the main features; the slightly oscillatory behaviour in higher order coefficients $\log (k)_1$ and $\log (k)_2$ can be neglected \textit{a posteriori}, since the fitting result reproduces accurately the input (see full lines vs. circles in Fig.~\ref{fig:fitrcomp} top panel).}
    \label{fig:fitMcomp}
\end{figure*}

\begin{table*}[htb]
    \caption{\label{tab:delayed-tau-fitting-params}
    Fitting coefficients of eqs.~\eqref{eq:gold-fitting-formula-a}--\eqref{eq:gold-fitting-formula-c} for the differential SFH sequences, eq.~\eqref{eq:gold-formula}, for SFHs from mass assembly (top) and \UniverseMachine (bottom).}
    \centering
    \begin{ruledtabular}
    \begin{tabular}{lrrrrrrr}
      & $\log(k)_{0j}$	& $\log(k)_{1j}$	& $\log(k)_{2j}$	& $\log(\tau)_{0j}$	& $\log(\tau)_{1j}$ &	$\log(t_f)_{0j}$ & $\log(t_f)_{1j}$ \\
        \hline\\[-3mm]
        $j=0$ & -5.161050	& -8.000000   & 8.000000    & 0.426175 & 1.182270	&  -1.794150	& -0.224799 \\
%$\log(\Mso)^1$
$j=1$ & 0.036759	& 1.789890 & -1.979550	& 0.001728	& -0.179328	& 0.497882	& 0.037822 \\
%$\log(\Mso)^2$
$j=2$ & 0.051082	& -0.095058	& 0.105380	& -0.005042	& 0.006703	& -0.021372	& -0.001594 \\ 
        \hline\\[-3mm]
        $j=0$ &	2.102330    & -8.000000 & 3.032460  & -7.788130 & 1.799550  & 1.024230  & -0.241818 \\
$j=1$ & -1.471980	& 1.780140	& -0.878934	& 1.713680	& -0.317082	& 0.030384	& 0.029203 \\
$j=2$ &	0.124825	& -0.093565	& 0.048557	& -0.088893	& 0.013846	& -0.001962	& -0.000735 \\  \\[-0.5mm]
    \end{tabular}
    \end{ruledtabular}
\end{table*}

%------------------------------------------------------
\subsubsection{Modelling}\label{sec:GRMA-modelling}
%Star formation 
Let $\Delta\Psi(t;r_1,r_2,\Mso)$ represent the SFH within physical galactocentric radii $r_1$ and $r_2$. 
In a first step, we write the galaxy-wide numerical scheme~\eqref{eq:dMsdt} as a parametric equation of the present-day stellar mass,
\begin{align}\label{eq:dMsdt-param}
    \frac{d\Mso}{dt}(t;\Mso) = \Psi(t;\Mso) - \int_{t_f}^t\frac{d f_{\rm ml}}{d\tau}\big|_{t-t'} \Psi(t';\Mso)\;\!dt'\,,
\end{align}
where $\Mso(t;\Mso)$ is the stellar mass history and $\Psi(t;\Mso)$ the SFH, both defined and determined in \S~\ref{sec:GMA}. Now, restricting eq.~\eqref{eq:dMsdt-param} to a radial shell within $r_1\to r_2$, we have
\begin{align}\label{eq:dMsdt-shell}
    \frac{d\Mso}{dt}(t;r_1,r_2,\Mso) = \Delta\Psi(t;r_1,r_2,\Mso) \nonumber \\
    - \int_{t_f}^t\frac{d f_{\rm ml}}{d\tau}\big|_{t-t'} \Delta\Psi(t';r_1,r_2,\Mso)\;\!dt'\,,
\end{align}
where $\Mso(r;,r_1,r_2,\Mso)$ is the integrated stellar mass history defined immediately after eq.~\eqref{eq:Mstar-r1-r2}, and is already known. 
Next, we solve eq.~\eqref{eq:dMsdt-shell} numerically by iteration, starting with the first guess $\Delta \Psi(t;r_1,r_2,\Mso) = d\Mso(t;r_1,r_2,\Mso)/dt$ in the mass loss integral. We found that 5 iterations are sufficient to attain convergence, but time resolution of at most $\lesssim 0.5$~Myr is necessary. The procedure has been tested on galaxy-wide SFHs $\Psi(t;\Mso)$---which are known---before inferring radial-shell SFHs. 
For convenience in analytical expressions, we also use the following differential notation (remembering that angles are already integrated out)
\begin{align}
    \frac{\pd\Psi}{\pd r}(t;r,\Mso) \equiv \lim_{\Delta r \to 0} \frac{\Delta\Psi(t;r\!-\!\Delta r/2,r\!+\!\Delta r/2,\Mso)}{\Delta r}\,.
\end{align}
In practice, we perform calculations on a grid of 13 finite radial zones with limits 
$\{$0., 0.05, 0.1, 0.2, 0.3, 0.4, 0.5, 0.75, 1., 1.5, 2., 3., 4., 5.$\}$
in units of \textit{present-day} projected Sérsic half-mass radii $R_{1/2}$. 
As for the galaxy-wide case, we fit the results with delayed-$\tau$ functions, finding, after inspection, that this parametric form~\eqref{eq:SFH-delayed-tau} holds for the differential history,
\begin{align}\label{eq:gold-formula}
    \frac{\pd \Psi}{\pd r}(t;r,\Mso) \simeq k\;\!\frac{t\!-\!t_f}{\tau}\;\!\exp\Big(\!\!-\!\frac{t\!-\!t_f}{\tau}\Big)\,,%\;M_{\odot}{\rm yr}^{-1}{\rm kpc}^{-1}\,,
\end{align}
for $t\geq t_f$, and $\pd\Psi/\pd r=0$ otherwise, and where $k$, $\tau$ and $t_f$ are fitting `constants' that now depend on $\Mso$ and $r$. 
For \UniverseMachine SFHs, delayed-$\tau$ fitting provides a better fitting than double power-law, therefore, we use eq.~\eqref{eq:gold-formula} for both SFH models. 
The result of the delayed-$\tau$ fitting is shown in Fig.~\ref{fig:fitrcomp}, where the small circles represent the obtained delayed-$\tau$ parameters for all 13 radial zones and a selection of five present-day stellar masses.

A close inspection of the radial behaviour of the delayed-$\tau$ parameters suggests the following parametric form (dashed lines in Fig.~\ref{fig:fitrcomp}):
\begin{align}
    \log(k) =&\; \sum_{i=0}^2 \log(k)_{i} \,\log(r)^i\,, \\
    \log(\tau) =&\; \sum_{i=0}^1 \log(\tau)_{i} \, r^i\,, \\
    \log(t_f) =&\; \sum_{i=0}^1  \log(t_f)_{i}\,r^i\,,
\end{align}
where $\log(k)_{i}$, $\log(\tau)_{i}$ and $\log(t_f)_i$ are fitting `constants' that depend only stellar mass, as shown in Fig.~\ref{fig:fitMcomp}.

A close inspection of the behaviour with stellar mass suggests second-order expansion in $\log(\Mso)$, such that the final parametric fitting formulae are (full lines in Fig.~\ref{fig:fitrcomp} and Fig.~\ref{fig:fitMcomp})
\begin{align}
    \log(k) =&\; \sum_{i=0}^2 \sum_{j=0}^2 \log(k)_{ij} \,\log(r)^i\;\!\log(\Mso)^j\,, \label{eq:gold-fitting-formula-a}\\
    \log(\tau) =&\; \sum_{i=0}^1 \sum_{j=0}^2 \log(\tau)_{ij}\, r^i \;\!\log(\Mso)^j \,, \label{eq:gold-fitting-formula-b}\\
    \log(t_f) =&\; \sum_{i=0}^1 \sum_{j=0}^2 \log(t_f)_{ij}\,r^i\;\!\log(\Mso)^j \,, \label{eq:gold-fitting-formula-c}
\end{align}
where $k$ is in units of $M_{\odot}$yr$^{-1}$kpc$^{-1}$, $t_f$ and $\tau$ are in Gyr, $r$ is in kpc and $\Mso$ is in solar masses. In total, we have $3\times 3+2\times 3+2\times 3=21$ coefficients which are reproduced in table \ref{tab:delayed-tau-fitting-params}.

Equation~\eqref{eq:gold-formula} together with the fitting expansions~\eqref{eq:gold-fitting-formula-a}--\eqref{eq:gold-fitting-formula-c} and fitting coefficients shown in table~\ref{tab:delayed-tau-fitting-params} constitute the main result of this work, which can now be applied to calculating rates of DM phenomenology in compact stars on cosmological scales (see, e.g., Ref.~\cite{2025arXiv250521256S}). 
Before doing so, let's test the formulae as much as possible.

\begin{figure*}
    \centering
    \includegraphics[width=0.45\linewidth]{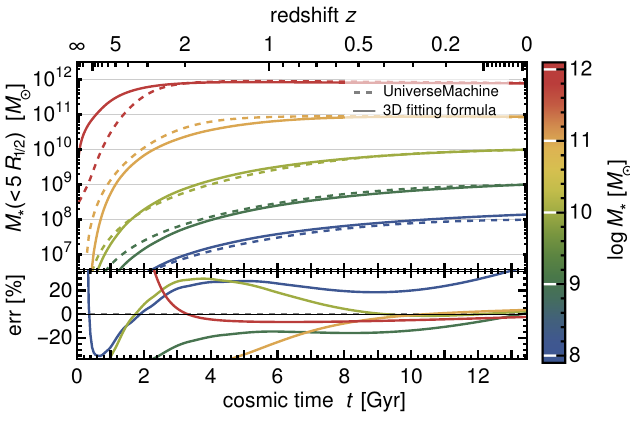}
    \includegraphics[width=0.45\linewidth]{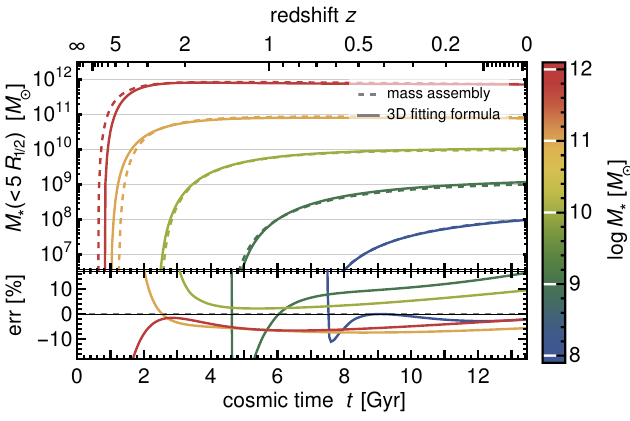}
    \caption{Total stellar mass histories within 5~$R_{1/2}$ as a function of cosmic time, as calculated by eq.~\eqref{eq:Mstar-r} (dashed lines), and compared to fitting formula \eqref{eq:gold-formula} (full lines), for a range of present-day stellar masses as indicated by colors. Left: \UniverseMachine\!\!. Right: mass assembly model. Mini bottom panels shows relative errors of the full vs. the dashed curves.}
    \label{fig:MtotalfitPsiMr}
\end{figure*}

\begin{figure*}
    \centering
    \includegraphics[width=0.45\linewidth]{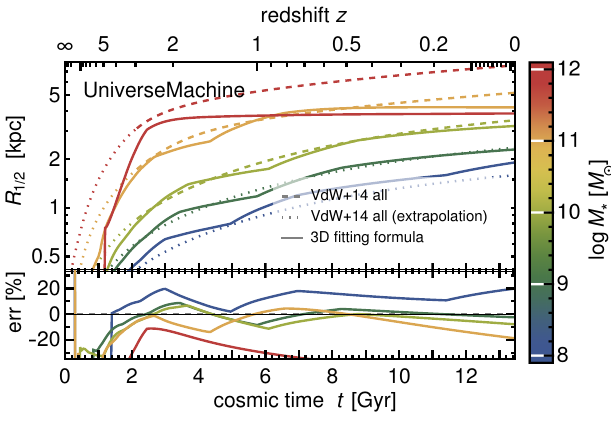}
    \includegraphics[width=0.45\linewidth]{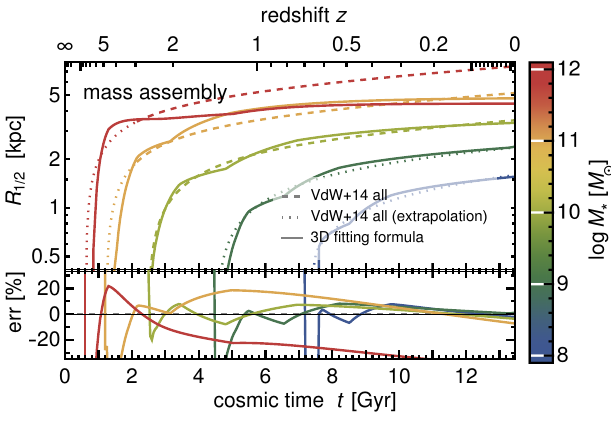}
    \caption{Comparison of circularized sky-projected half-mass sizes of simulated galaxies with fitting formula eq.~\eqref{eq:gold-formula} (full lines) compared to the input scaling relation $R_{1/2}(t;\Mso)$ as defined in \S~\ref{sec:galaxy-structure} when assuming mass histories of \S~\ref{sec:GMA}     (dashed and dotted lines), for a selection of present-day stellar masses. The dashed lines show the region of actual validity of eq.~\eqref{eq:circularized-half-light-radius}, i.e. where observational data was used by %Van der Wel et al. (2014)
    Ref.~\protect\cite{2014ApJ...788...28Vmax10}, dotted lines are extrapolations. Left: \UniverseMachine. Right: mass assembly model. Mini bottom panels shows relative errors of the full vs. the dashed/dotted curves.}
    \label{fig:RefitPsiMr}
\end{figure*}

\begin{figure*}[htb]
    \centering
    \includegraphics[width=0.45\linewidth]{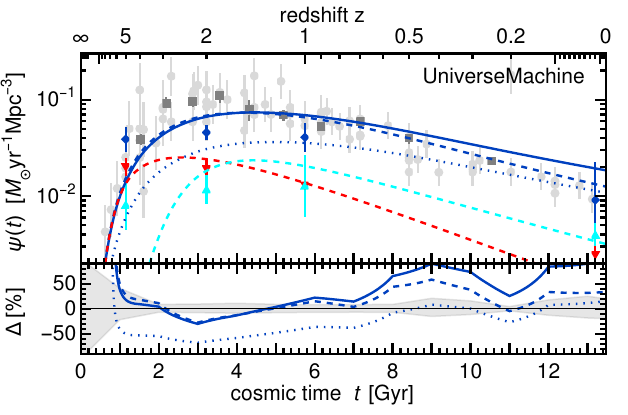}
    \includegraphics[width=0.54\linewidth]{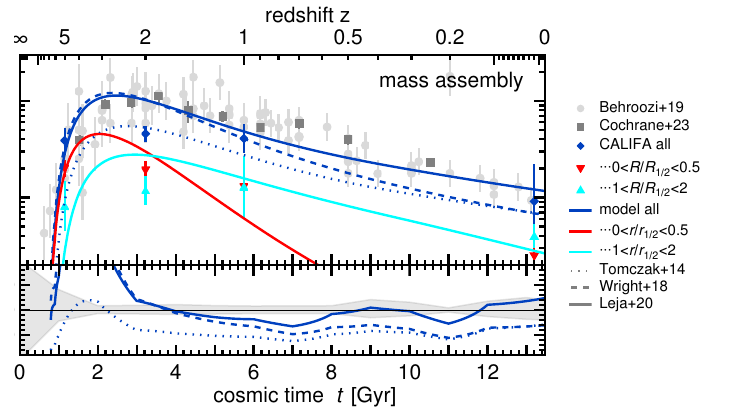}
    \caption{Same as Fig.~\ref{fig:SFRD} but for the differential fitting formulae, i.e. eq.~\eqref{eq:gold-formula} with eqs.~\eqref{eq:gold-fitting-formula-a}--\eqref{eq:gold-fitting-formula-c} and coefficients from table~\ref{tab:delayed-tau-fitting-params}, integrated according to eq.~\eqref{eq:SFRD-fitting}, for \UniverseMachine SFHs (left) and mass assembly SFHs (right).
    Also shown is $\psi(t)$ limited to galactocentric \textit{three-dimensional} radial shells from fitting formula (red and cyan lines) compared to $\psi(t)$ within \textit{sky-projected} radial shells from the CALIFA survey %(Lopez-Fernandez (2018)
    \protect\cite{2018A&A...615A..27L} (red and cyan triangles).  The mini bottom panels show relative errors (same line styles) with respect to the %LOFAR%(Cochrane et al. 2023)\cite{2023MNRAS.523.6082C}
    data mean \protect\cite{2019MNRAS.488.3143B,2023MNRAS.523.6082C} and compared to its 1$\sigma$ confidence region (gray shaded area).} 
    \label{fig:SFRDinout}
\end{figure*}

%------------------------------------------------------
\subsubsection{Consistency testing}
\label{sec:SFH-testing}
Since, to our knowledge, SFH models in galactocentric radial shells are still missing in the literature,
testing $\pd\Psi/\pd r$ consists here in trivially reproducing the input physics. Nevertheless, we found spatially resolved SFRD determinations from the CALIFA survey %Lopez-Fernandez (2018) 
\cite{2018A&A...615A..27L} and these constitute an additional, independent, though qualitative, test.

Test number one consists in reproducing the input mass histories $\Mso(t;\Mso)$ when summed over radial shells. Given that we limited the fitting procedure to five half-mass radii, we compare the integrated mass histories up to $5~R_{1/2}$ (see Fig.~\ref{fig:MtotalfitPsiMr}). For the mass assembly model, comparison is globally satisfying, with a relative error of at most 10\% for most redshifts, reflecting the fidelity of the fitting scheme, eqs.~\eqref{eq:gold-fitting-formula-a}--\eqref{eq:gold-fitting-formula-c}. 
For \UniverseMachine\!\!, the error is less than 30\% for most parts. This difference in relative error is probably related to the fact that the delayed-$\tau$ fitting is not as optimal for \UniverseMachine SFHs as it is for mass assembly SFHs.

Test number two consists in reproducing the input galaxy size histories $R_{1/2}(t;\Mso)$. This can be done by solving numerically the condition that the mass up to a certain radius be equal to half the integrated mass at that time. The results are shown in Fig.~\ref{fig:RefitPsiMr}. For both SFH models, the relative error remains below 20\%  all relevant redshifts and stellar masses. The only exception are ultra-massive galaxies ($\Mso=10^{12}M_{\odot}$) with up to 40\% relative error. For global transient volumetric rates, this shouldn't be much of a problem because the contribution of ultra-massive galaxies to the SFRD is negligible. However, when it comes to predictions for the most outer part of the host-offset distribution (fig.~1d in Ref.~\cite{2025arXiv250521256S}), we expect to underestimate the true rate, especially at low redshifts, because these model ultra-massive galaxies are more compact than the input scaling relations. 

Test number three (and maybe the most important) consists in reproducing the SFRD when integrated over radial shells and the galaxy stellar mass function, 
\begin{align}\label{eq:SFRD-fitting}
    \psi(t) = \iint \frac{\pd\Psi}{\pd r}(t;r,\Mso)\;\!\phi(\Mso)\;\!dr\;\!d\Mso\,.
\end{align}
The results are shown in Fig.~\ref{fig:SFRDinout} compared to the data already shown in Fig.~\ref{fig:SFRD}. The quality of the fittings is comparable to the point-like case. For the mass assembly SFHs, using the galaxy stellar mass function of Ref.~\cite{2020ApJ...893..111L} (full line) still provides the better match to the data, with relative error less than 30\% for most redshifts, except for a short period between $2\lesssim z \lesssim 5$ where it becomes a factor of two higher. 
For the \UniverseMachine SFHs, using the galaxy stellar mass function of Ref.~\cite{2018MNRAS.480.3491W} (dashed line) now provides a slightly better fit than using that of Ref.~\cite{2014ApJ...783...85T} (dotted line), with relative error better than 30\% overall, except locally around $z\sim 0.4$ with up to 50\%.

The final test consists in comparing predictions of fitting formula~\eqref{eq:gold-formula} with spatially resolved SFRD data from the CALIFA survey %Lopez-Fernandez (2018) 
\cite{2018A&A...615A..27L}. 
The red and cyan curves in Fig.~\ref{fig:SFRDinout} show the predicted SFRD within \textit{three-dimensional} galactocentric spheres with $r < 0.5~R_{1/2}$ and shells with $r \in [R_{1/2},\,2~R_{1/2}]$, respectively, while red down and cyan up triangles show the observed SFRD within \textit{sky-projected} galactocentric `cylinders' with $R < 0.5~R_{1/2}$ and `donuts' with $R \in [R_{1/2},\,2R_{1/2}]$, respectively. Because of the projection effect on the observational data, the comparison has to remain on a qualitative level: as can be seen, both predictions and data points illustrate the well-known fact (usually referred to as inside-out formation) that stars form earlier in galaxy centers and later in their peripheries. 
The effect is naturally less pronounced in the observational data than in predictions because of the mixing of radial shells in `cylinders' or 'donuts' oriented along the line of sight.

%%%%%%%%%%%%%%%%%%%%%%%%%%%%%%%%%%%%%%%%%%%%%%%%%%%%%%%%%
%%%%%%%%%%%%%%%%%%%%%%%%%%%%%%%%%%%%%%%%%%%%%%%%%%%%%%%%%
\section{Discussion}\label{sec:discussion}

This paper is dedicated to the construction of average SFHs in galactocentric radial shells, in order to estimate DM-induced transient rates in compact stars. The basis of the framework are empirical galaxy structure relations and two different models of average SFHs (a simple mass assembly model and \UniverseMachine simulation results). 
Combining structure and evolution allowed us to simulate SFHs in galactocentric radial zones, up to five half-mass radii. 
Furthermore, we have derived closed-form fitting expressions, eq.~\eqref{eq:gold-formula}, accurate with respect to input physics to within a few tens of per cent for both SFH models. 
These are useful for numerical simulation procedures to quickly explore the parameter spaces of DM candidate theories.

In our companion paper, Ref.~\cite{2025arXiv250521256S}, we apply the framework to the specific situation where encounters between WDs and PBHs trigger normal SNe Ia \cite{2021PhRvL.127a1101S}. To this end, we have provided, in \hyperref[sec:WD-formation]{Appendix}, a supplementary framework to efficiently estimate the delayed WD mass function, based on simple scaling relations of stellar IMF,  initial-final mass relation, and main-sequence life times. Possible effects of metallicity and SFR on the IMF and the initial-final mass relation are also considered, however, these should be treated with caution, because the choice of the IMF has influences on the SFH [via eq.~\eqref{eq:mass-loss-rate}]. To our knowledge, SFH simulations such as UniverseMachine but assuming a non-canonical IMF, are still missing in the literature.
Finally, the present model of WD formation does not yet include metallicity gradients with galactocentric radial offset.

%%%%%%%%%%%%%%%%%%%%%%%%%%%%%%%%%%%%%%%%%%%%%%%%%%%%%%%%%
%%%%%%%%%%%%%%%%%%%%%%%%%%%%%%%%%%%%%%%%%%%%%%%%%%%%%%%%%
\begin{acknowledgments}
The author is thankful to Silvia Toonen and Peter Behroozi for useful correspondence and discussions and to the anonymous referees for their comments and suggestions. The work was initiated with financial support of FAPES/CAPES DCR Grant No. 009/2014. 
\end{acknowledgments}

%%%%%%%%%%%%%%%%%%%%%%%%%%%%%%%%%%%%%%%%%%%%%%%%%%%%%%%%%
%%%%%%%%%%%%%%%%%%%%%%%%%%%%%%%%%%%%%%%%%%%%%%%%%%%%%%%%%
\appendix*

%%%%%%%%%%%%%%%%%%%%%%%%%%%%%%%%%%%%%%%%%%%%%%%%%%%%%%%%%
%%%%%%%%%%%%%%%%%%%%%%%%%%%%%%%%%%%%%%%%%%%%%%%%%%%%%%%%%
\appsection{Remnant formation and mass function}\label{sec:WD-formation}

Stellar remnant formation is delayed with respect to star formation by the main-sequence life time and possibly by merger delay times. Here we focus on the formation of WDs, and in particular on the range of masses of interest for sub-Chandrasekhar SNe Ia, $0.9M_{\odot}\lesssim m_w\lesssim 1.1M_{\odot}$ %(Sim et al. 2010; Blondin et al. 2017; Shen et al. 2018, 2021a)
\cite{2010ApJ...714L..52S,*2017MNRAS.470..157B,*2018ApJ...854...52S,*2021ApJ...909L..18S}. The procedure can be readily adapted to other mass ranges or to the formation of NSs.

%===========================================================
\subsection{Stellar mass function}\label{sec:IMF}
There are strong theoretical and empirical reasons to believe that the stellar IMF is not universal, but depends on gas phase metallicity ($Z$), mainly at lower stellar mass, $m_{\star}\lesssim 1M_{\odot}$ %(Marks et al. 2012; Yan et al. 2020)
\cite{2012MNRAS.422.2246M,2020A&A...637A..68Y}, and on SFR ($\Psi$) at higher stellar mass, $m_{\star}\gtrsim 1M_{\odot}$ %(Kroupa 2003...; Weidner et al. 2013; Fontanot et al. 2017; ...)
\cite{2003ApJ...598.1076K,2013MNRAS.436.3309W,2017MNRAS.464.3812F}. 

We construct a galaxy-wide IMF with $Z$-dependent slope for $m_{\star}<1M_{\odot}$ as given by eq.~9 of %Yan et al. (2020)
Ref.~\cite{2020A&A...637A..68Y}, which is 
consistent with %Marks et al. (2012)
Ref.~\cite{2012MNRAS.422.2246M}, and $\Psi$-dependent slope for $m_{\star}>1M_{\odot}$ based on data provided by %Fontanot et al. (2017)
Ref.~\cite{2017MNRAS.464.3812F}, and consistent with the work %Weidner et al. (2013), 
of Ref.~\cite{2013MNRAS.436.3309W}. We neglect the very marginal $Z$-dependence at higher stellar mass %Fontanot et al. (2017)
\cite{2017MNRAS.464.3812F}. Let $\xi_{\star}(m_{\star})dm_{\star}$ be the number of main-sequence stars formed in a single burst in the mass range $m_{\star}\to m_{\star}\!+\!dm_{\star}$. We use a four-slope parametric form
\begin{align}\label{eq:IGIMF}
    \xi_{\star}(m_{\star}) = \begin{cases}
        0 & m_{\star}<m_{\rm min} \\
        k_1 (m_{\star}/m_{\rm min})^{-\alpha_1} & m_{\rm min}\leq m_{\star} < m_0 \\
        k_1\;\! k_2 \;\!(m_{\star}/m_0)^{-\alpha_2} & m_0\leq m_{\star} < m_1 \\
        k_1\;\! k_2\;\!k_3\;\! (m_{\star}/m_1)^{-\alpha_3} & m_1\leq m_{\star} < m_{\rm br} \\
        k_1\;\! k_2\;\!k_3\;\!k_4\;\! (m_{\star}/m_{\rm br})^{-\alpha_4} & m_{\rm br}\leq m_{\star} < m_{\rm max} \\
        0 & m_{\star}\geq m_{\rm max}
    \end{cases}\,,
\end{align}
where the normalization factors are 
\begin{align}
    k_1 = &\; c_1\log(\Psi)^3 + (c_2\;\!Z\!+\!c_3)\log(\Psi)^2 + (c_4\;\!Z^2\!+\!c_5\;\!Z\!+\!c_6)\nonumber\\
     &\;  \times \log(\Psi) + c_7\;\!Z^2 + c_8\;\!Z+c_9 \label{eq:k-IGIMF} \\
    k_2 =&\; (m_0/m_{\rm min})^{-\alpha_1} \\
    k_3= &\; (m_1/m_0)^{-\alpha_2} \\
    k_4= &\; (m_{\rm br}/m_1)^{-\alpha_3} 
\end{align}
with $c_1=-0.0008513$, $c_2=-0.003655$, $c_3=-0.008166$, $c_4=56.0302$, $c_5=-2.4958$, $c_6=-0.05287$, $c_7=2820.93$, $c_8=157.79$, $c_9=2.507$; and where the masses are  
\begin{align}
    m_{\rm min} =&\; 0.08 \\
    m_0 = &\; 0.5 \\
    m_1 =&\; \begin{cases}
        1.287 & \log(\Psi)\leq -0.5 \\
        1.0 & \log(\Psi)>-0.5
    \end{cases} \\
    m_{\rm br} =&\; \begin{cases}
        {\rm min}(10^{c_1\log(\Psi)+c_2},67) & \log(\Psi)\leq -0.5 \\
        10^{c_3\log(\Psi) + c_4} & \log(\Psi) > -0.5
    \end{cases} \\
    m_{\rm max} =&\; {\rm min}(10^{c_5\log(\Psi)+c_6},100)
\end{align}
with $c_1=0.442$, $c_2=2.588$, $c_3=-0.0324$, $c_4=1.0774$; $c_5=0.472$, $c_6=2.812$; 
and where the slopes are
\begin{align}
    \alpha_1 = &\; 1.3+35\;\!(Z-0.02) \\
    \alpha_2 = &\; 2.3+35\;\!(Z-0.02) \\
    \alpha_3 = &\; 10^{c_1\log(\Psi)+c_2} \label{eq:IMF-alpha-3} \\
    \alpha_4 = &\; 10^{c_3\log(\Psi)^2+c_4\log(\Psi)+c_5}
\end{align}
with $c_1=-0.0398$, $c_2=0.3862$, $c_3=0.0137$, $c_4=-0.1173$, $c_5=0.4250$. 

\begin{figure}[b]
    \centering
    \includegraphics[width=\linewidth]{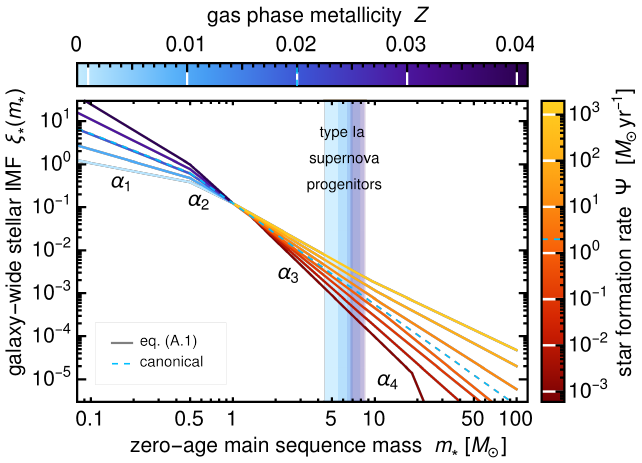}
    \caption{Galaxy-wide stellar IMF with metallicity-dependent low-mass slope \cite{2020A&A...637A..68Y} (blue colors)
    %(Yan et al. 2020)
    and SFR-dependent high-mass slope %(Fontanot et al. 2017)
    \cite{2017MNRAS.464.3812F} (yellow-red colors) given by eq.~\eqref{eq:IGIMF} and compared the canonical stellar IMF %(Kroupa 2001, Kroupa et al. 2013)
    \protect\cite{2001MNRAS.322..231K,*2013pss5.book..115K} (cyan dashed line).
    %(Kroupa 2001, green dashed line) 
    The normalization has been set to match the canonical at $m_{\star}=1M_{\odot}$ for illustrative purposes. The vertical shaded stripes indicate the SN Ia progenitor mass range for different metallicities (reflected by blue tones) and were calculated using eq.~\eqref{eq:IFMR-Umeda+99} and assuming $m_w\in [0.9,\,1.1]M_{\odot}$.}
    \label{fig:IGIMF}
\end{figure}

The result is shown in Fig.~\ref{fig:IGIMF} and compared to the canonical IMF %(Kroupa 2001, Kroupa et al. 2013)
\cite{2001MNRAS.322..231K,*2013pss5.book..115K}, which coincides with the Milky-Way thin disc conditions ($Z=0.02$, $\Psi=2M_{\odot}$yr$^{-1}$). For the purpose of that figure and a better understanding of the slopes, the overall normalization factor there is different from that of eq.~\eqref{eq:k-IGIMF}
\begin{align*}
    k_1 = 0.8387\times10^{38.92 Z+0.126}\,,
\end{align*}
such that all curves shown have the value of the canonical IMF at $m_{\star}=1M_{\odot}$.

Since the range of zero-age main sequence masses that potentially explode as SNe Ia, roughly $4M_{\odot}\lesssim m_{\star}\lesssim 8M_{\odot}$ (see Fig.~\ref{fig:IGIMF}) depending on metallicity (see \S~\ref{sec:IFMR}), lies within the interval where the IMF slope depends on SFR [see Fig.~\ref{fig:IGIMF} and eq.~\eqref{eq:IMF-alpha-3}], the SN Ia progenitor slope will depend on $\Psi$ accordingly. Conversely, lower metallicity means less low-mass stars, and hence more SN Ia progenitors, for a same amount of stars formed.

\begin{figure}
    \centering
    \includegraphics[width=\linewidth]{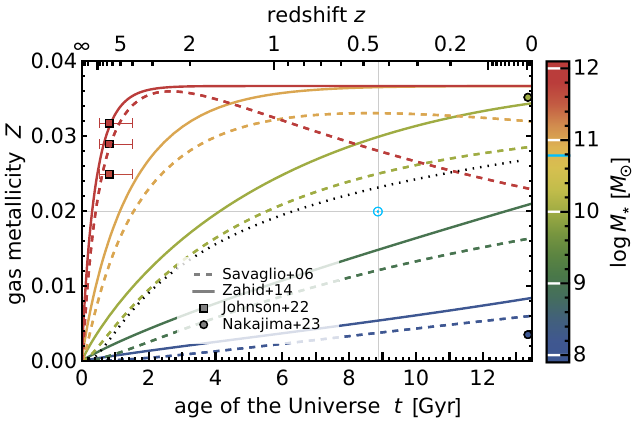}
    % produced with stellar-mass-functions.nb
    \caption{Mean gas phase metallicity evolution as a function of cosmic time for present-day host stellar masses  as indicated by colors according to fitting relations \protect\cite{2006NJPh....8..195S,2014ApJ...791..130Z}, and recent determinations %Johnson (2022), Nakajima 2023
    \protect\cite{2023MNRAS.526.5911J,2023ApJS..269...33N}.  The black dotted line indicates the average metallicity of star-forming galaxies \cite{2008MNRAS.388.1487L}.
    The formation time and metallicity of the Sun %(von Steiger \& Zurbuchen 2016)
    \protect\cite{2016ApJ...816...13V}, and the Milky-Way stellar mass \protect\cite{2015ApJ...806...96L}
    are shown in cyan for comparison.}
    \label{fig:Zoft}
\end{figure}

Let $\xi_{\star}(t;m_{\star},M_{\star})$ be the stellar IMF history for the metallicity and SFR conditions in a galaxy with present-day stellar mass $M_{\star}$ at cosmic time $t$. We already know the SFR conditions from the SFH models $\Psi(t;M_{\star})$ of \S~\ref{sec:GMA}. Now, let $Z(t;M_{\star})$ be the average gas phase metallicity for given present-day stellar mass and cosmic time. In Fig.~\ref{fig:Zoft}, we compare some recent determinations of $Z(t;M_{\star})$. For choices of $Z(t;M_{\star})$ and $\Psi(t;M_{\star})$, one may determine easily $\xi_{\star}(t;m_{\star},M_{\star})$.
Thus, the formation history of stars with mass in the range $m_{\star}\to m_{\star}\!+\!dm_{\star}$ in a galaxy with present-day stellar mass $M_{\star}$, is given by 
\begin{align}\label{eq:SFH-mass-range}
    \Psi(t;M_{\star})\;\!\xi_{\star}(t;m_{\star},M_{\star})\;\!dm_{\star}\,,
\end{align}
and the same but restricted to a galactocentric radial shell within $r\to r\!+\!dr$ is given by  
\begin{align}\label{eq:SFH-mass-range-shell}
    \frac{\pd \Psi}{\pd r}(t;r,M_{\star})\;\!\xi_{\star}(t;m_{\star},M_{\star})\;\!dm_{\star}\,dr\;\!.
\end{align}

\begin{figure}[htb]
    \centering
    \includegraphics[width=\linewidth]{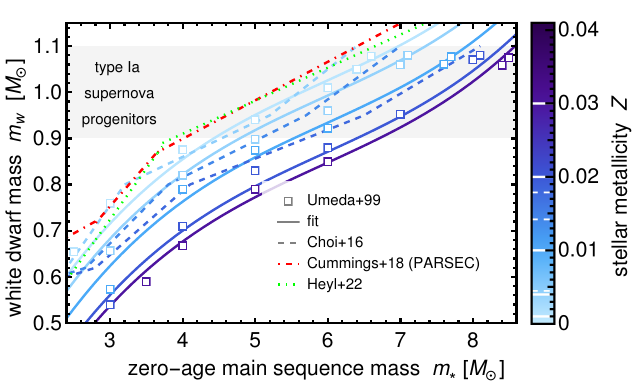}
    \caption{Initial-final mass relation, i.e. WD mass as a function of zero-age main sequence mass for various stellar metallicities according to numerical simulations %Umeda et al. (1999)
    of Ref.~\protect\cite{1999ApJ...513..861U} (blue squares) and fitted by eq.~\eqref{eq:IFMR-Umeda+99} (flue full lines) and compared to numerical simulations of Ref.~\protect\cite{2016ApJ...823..102C} (blue dashed lines) and empirical results of %Cummings (2018)
    Ref.~\cite{2018ApJ...866...21C} (red dot-dashed line) and 
    %Heyl et al. (2022)
    Ref.~\cite{2022ApJ...926..132H} (green dotted line). The light gray shaded stripe indicates the sub-Chandrasekhar SN Ia progenitor mass range.}
    \label{fig:IFMR}
\end{figure}

%===========================================================
\subsection{Initial-final mass relation}\label{sec:IFMR}
Metallicity also influences stellar evolution and the final remnant mass for a given zero-age main-sequence mass %(Umeda et al. 1999, Choi et al. 2016)
\cite{1999ApJ...513..861U,2016ApJ...823..102C}.
In Fig.~\ref{fig:IFMR}, we compare initial-final mass relations (IFMRs) for non-rotational stellar simulations with various metallicities %(Umeda et al. 1999, Choi et al. 2016)
\cite{1999ApJ...513..861U,2016ApJ...823..102C} with empirical IFMRs %Cummings (2018),  %Heyl et al. (2022)
\cite{2018ApJ...866...21C,2022ApJ...926..132H}.
In this work, we adopt the results of Ref.~\cite{1999ApJ...513..861U}, because these cover the largest range of metallicities and are still consistent with Ref.~\cite{2016ApJ...823..102C}. Observed WD masses are usually found slightly higher offset with respect to theoretical predictions \cite{2018ApJ...866...21C} (see Fig.~\ref{fig:IFMR}). For convenience, we fit the results of %(Umeda et al. 1999)
Ref.~\cite{1999ApJ...513..861U} assuming that extrapolation beyond $Z>0.03$ (up to $0.035$) holds,
\begin{align}\label{eq:IFMR-Umeda+99}
    m_w = c_1\;\!m_{\star}^3+c_2\;\!m_{\star}^2+c_3\;\!m_{\star} + c_4\;\!{\rm erf}(c_5\;\!Z)+c_6\,,
\end{align}
where ${\rm erf}(x)=(2/\sqrt{\pi})\int_0^x \exp(-t^2)dt$ is the error function and $c_1=0.003689$, $c_2=-0.06585$, $c_3=0.4637$, $c_4=-0.1956$, $c_5=50.6603$, $c_6=-0.1714$. The result is shown as full blue lines in Fig.~\ref{fig:IFMR}. 
The following alternative fitting is more precise in the data range, but extrapolation to $Z> 0.03$ leads to inversion of the trends,
\begin{align}\label{eq:IFMR-Umeda+99-old}
    m_w =&\; c_1\;\!m_{\star}^3+(c_2\;\!Z+c_3)\;\!m_{\star}^2+(c_4\;\!Z^2+c_5\;\!Z+c_6)\;\!m_{\star} \nonumber \\
    &\;+ c_7\;\!Z^2+c_8\;\!Z+c_9\,,
\end{align}
where $c_1=0.00625$, $c_2=-0.520$, $c_3=-0.101$, $c_4=-21.466$, $c_5=6.591$, $c_6=0.607$, $c_7=323.772$, $c_8=-31.958$, $c_8=-0.325$.
As we shall see, using either of the fittings results in only small differences on the WD mass function.  As for the case of the IMF, let us call $m_w(t;m_{\star},M_{\star})$ the IFMR history for the  environmental conditions at cosmic time $t$ in a galaxy with present-day stellar mass $M_{\star}$, and $m_{\star}(t;m_w,M_{\star})$ it's inverse. 

%===========================================================
\subsection{White dwarf mass function}\label{sec:WD-mass-function}
Despite the monumental progress with \textit{Gaia} space craft, the current empirical knowledge of the WD mass function stems from a still relatively small population (compared to main-sequence stars) within a few hundreds of parsecs distance from the Sun % Kilic 
(e.g., Ref.~\cite{2020ApJ...898...84K}). 
Binary population synthesis (BPS) calculations have reproduced the observed record for Solar-system conditions of the stellar IMF and metallicity
\cite{2020ApJ...898...84K,2020A&A...636A..31T}. 
Interestingly, it is found that even though a large fraction ($\sim 35\%$) of all WDs form as product of a binary merger, the shape of the WD mass function does not change significantly due to mergers \cite{2020A&A...636A..31T}, and the effect is particularly negligible in the specific mass range of SN Ia progenitors.
Therefore, to a very good approximation, the WD mass function may be estimated simply as the mathematical pull-back of the stellar IMF by the inverse of the IFMR. This will allows us to make efficient predictions for the different environmental conditions (metallicity and SFR) across the entire Universe without recurring to BPS calculations. Implicitly, this supposes that the BPS results of non-varying WD mass function due to mergers apply also to all environmental conditions.

Let $\xi_w(m_w)\;\!dm_w$ be the number of WDs with mass in the range $m_w\to m_w\!+\!dm_w$ per solar mass of stars formed. After completion of WD formation delay times, we have
\begin{align}\label{eq:xiwmw}
    \xi_w(m_w)\;\!dm_w = \frac{\xi_{\star}[m_{\star}(m_w)]\;\!dm_{\star}%\;\!f_{\rm CO}(m_w)
    }{\int_{0.08M_{\odot}}^{125M_{\odot}} m_{\star}\;\!\xi_{\star}(m_{\star})\;\!dm_{\star}}\,,
\end{align}
where $m_{\star}(m_w)$ is the inverse of the IFMR, and the denominator calculates the average stellar mass for the IMF under consideration.

In case of non-Universal IMF and IFMR, we also have the WD mass function for stars formed under the conditions at cosmic time $t$ in a galaxy with present-day stellar mass $M_{\star}$, and observed after remnant formation delay times are completed,
\begin{align}\label{eq:xiwtmwMs}
    \xi_w(t;m_w,M_{\star})\;\!dm_w = \frac{\xi_{\star}[t;m_{\star}(t;m_w,M_{\star}),M_{\star}]\;\!dm_{\star}%\;\!f_{\rm CO}(m_w)
    }{\int_{0.08M_{\odot}}^{125M_{\odot}} m_{\star}\;\!\xi_{\star}(t;m_{\star},M_{\star})\;\!dm_{\star}}\,.
\end{align}

If the remnant formation delay times were zero, we would have the WD formation history in analogy with eq.~\eqref{eq:SFH-mass-range},
\begin{align}\label{eq:WD-formation-history-mass-range}
    \Psi(t;M_{\star})\;\!\xi_w(t;m_w,M_{\star})\;\!dm_w\,.
\end{align}
In order to account for the remnant formation delay times, let us introduce $\Phi_w(\tau;m_w)$ the WD formation delay time distribution for given WD mass following a star formation burst at $\tau=0$. The actual WD formation history, taking into account remnant formation delay times, is the convolution of the WD formation delay time distribution with the `bare' WD formation history, eq.~\eqref{eq:WD-formation-history-mass-range},
\begin{align}\label{eq:f-mw-rm-Mstar-0}
     \!\int_0^t\!\Phi_w(\tau;m_w)\;\!\Psi(t\!-\!\tau;M_{\star})\;\!\xi_w(t\!-\!\tau;m_w,M_{\star})\;\!d\tau\,dm_w\,,
\end{align}
which represents the number of WDs formed with mass in the range $m_w\to m_w\!+\!dm_w$ at cosmic time $t$ in a galaxy with present-day stellar mass $M_{\star}$. 
Restricting this expression to a radial galactocentric shell, we define
\begin{widetext}
\begin{align}\label{eq:f-mw-rm-Mstar}
    f(t;m_w,r,M_{\star}) \;\!dm_w\;\!dr \equiv \!\int_0^t\!\!\!\Phi_w(\tau;m_w)\;\!\frac{\pd \Psi}{\pd r}(t\!-\!\tau;r,M_{\star})\;\!\xi_w(t\!-\!\tau;m_w,M_{\star})\;\!d\tau\;\!dm_w\;\!dr,
\end{align}
\end{widetext}
which represents the number of WDs formed in the mass range $m_w\to m_w\!+\!dm_w$ in a galactocentric radial shell within $r\to r\!+\!dr$ in a galaxy with present-day stellar mass $M_{\star}$ at cosmic time $t$. Equation~\eqref{eq:f-mw-rm-Mstar} represents the second main result of this work, and is the starting point for calculating WD-PBH collision rates in our companion paper, Ref.~\cite{2025arXiv250521256S}.

The WD formation delay time from single stellar evolution is approximately equal to their main-sequence lifetime, $\tau_{\star} \simeq 7~$Gyr$~(m_{\star}/M_{\odot})^{-2.5}$ %(Fontaine 2001; Choi et al. 2016)
\cite{2001PASP..113..409F,2016ApJ...823..102C}, while that from binary mergers depends on main-sequence masses and initial separation, but is on average four times that from single stellar evolution % Temmink
\cite{2020A&A...636A..31T}. 
% NOT (Toonen et al. 2017) \cite{2017A&A...602A..16T} 
Thus, in order to keep things simple, we may set
\begin{align}
    \Phi_w(\tau; m_w) \simeq 0.65\;\!\delta(\tau\!-\!\tau_{\star})+0.35\;\!\delta(\tau\!-\!4\tau_{\star})\,,
\end{align}
where $\delta(t)$ is Dirac's delta function. We note that the remnant formation delay times in the SN Ia progenitor mass range are relatively short compared to cosmic time scales.

The actual WD population can be calculated by time-integrating eq.~\eqref{eq:f-mw-rm-Mstar},
\begin{align}\label{eq:local-WD-mass-function}
    \int_0^t\!f(t';m_w,r,M_{\star})\;\!dt'\,dm_w\;\!dr\,.
\end{align}
which represents the actual number of WDs with mass in the range $m_w\to m_w\!+\!dm_w$  present at cosmic time $t$ in a galactocentric radial shell within $r\to r\!+\! dr$ in a galaxy With present-day stellar mass. 
Thus, we may estimate the local, near Solar-System WD population by setting $t\simeq 13.5$~Gyr, $r\simeq 8$~kpc, and $M_{\star}\simeq 10^{10}M_{\odot}$ in eq.~\eqref{eq:local-WD-mass-function}.

\begin{figure}[htb!]
    \centering
    \includegraphics[width=\linewidth]{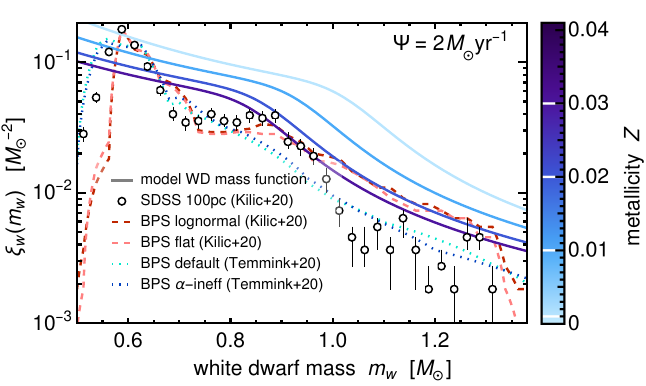}
    \caption{Differential number of WDs per solar mass of stars formed as a function of mass and for a selection of gas-phase metallicities as indicated in the bar chart (blue full lines) and as calculated  using the empirical metallicity-dependent IMF of %Yan et al. (2020)
    Ref.~\protect\cite{2020A&A...637A..68Y} and the fitting of eq.~\eqref{eq:IFMR-Umeda+99} to the metallicity-dependent IFMR of %Umeda et al. (1999)
    Ref.~\protect\cite{1999ApJ...513..861U}. The result is compared to local thin-disc 100~pc volume- and magnitude-limited observational data from SDSS (black circles, \cite{2020ApJ...898...84K}) and Solar-metallicity ($Z=0.02$) BPS simulation results of %Kilic, Temmink 
    Refs.~\cite{2020ApJ...898...84K,2020A&A...636A..31T} (line styles as indicated); these literature results were normalized by the number of WDs per Solar mass of stars formed in the thin disc, which is 0.032, assuming $Z=0.02$ and range $m_w \in [0.56,1.38]M_{\odot}$, corresponding to formation timescales $\tau_{\star}<10$~Gyr.}
    \label{fig:xiwmw}
\end{figure}

In Fig.~\ref{fig:xiwmw}, we show the differential number of WDs per solar mass of stars formed 
%given by eq.~\eqref{eq:xiwtmwMs} 
as a function of WD mass and for a selection of metallicities, under assumption of the empirical $Z$-dependent stellar IMF of %Yan et al. (2020)
Ref.~\cite{2020A&A...637A..68Y} and the fitting of eq.~\eqref{eq:IFMR-Umeda+99} to the $Z$-dependent IFMR of %Umeda et al. (1999)
Ref.~\cite{1999ApJ...513..861U} (blue lines). The top-part of the stellar IMF is assumed canonical ($\Psi=2M_{\odot}$yr$^{-1}$), and we neglect remnant formation delay times for the purpose of that figure. As can be seen, lower metallicity generates more WDs per solar mass of stars formed. The WD mass function slope does not vary significantly with metallicity. 
Also shown in Fig.~\ref{fig:xiwmw} are the local thin-disc empirical \cite{2020ApJ...898...84K}) and simulated %Kilic, Temmink 
\cite{2020ApJ...898...84K,2020A&A...636A..31T} WD mass functions in units of WDs per solar mass formed.
Broad agreement is achieved for the SN Ia progenitor mass range ($0.9M_{\odot}\lesssim m_w\lesssim 1.1M_{\odot}$) both in slope and normalization for solar metallicity ($Z=0.02$).

\begin{figure}[htb]
    \centering
    \includegraphics[width=\linewidth]{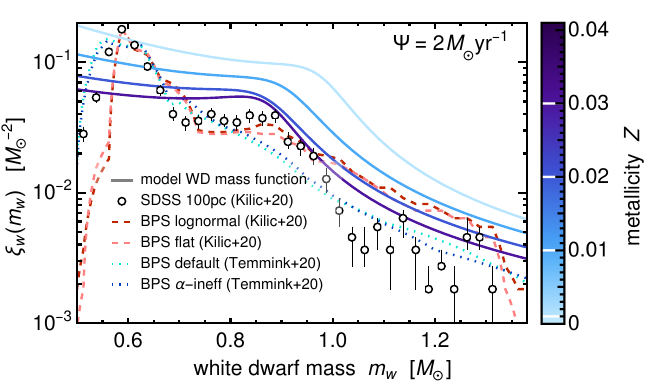}
    \caption{Same as Fig.~\ref{fig:xiwmw}, but using the IFMR fitting of eq.~\eqref{eq:IFMR-Umeda+99-old} instead of eq.~\eqref{eq:IFMR-Umeda+99}.}
    \label{fig:xiwmw-old}
\end{figure}
In Fig.~\ref{fig:xiwmw-old}, we show the WD mass function for the alternative fitting of eq.~\eqref{eq:IFMR-Umeda+99-old} instead of eq.~\eqref{eq:IFMR-Umeda+99}. The difference is a more pronounced bump around $m_w \sim 0.8M_{\odot}$. The difference for higher-mass ($m_w\gtrsim 0.9M_{\odot}$) WDs is negligible, such that either fitting formulae are equally suited for studies involving SNe Ia.

\begin{figure}[htb]
    \centering
    \includegraphics[width=\linewidth]{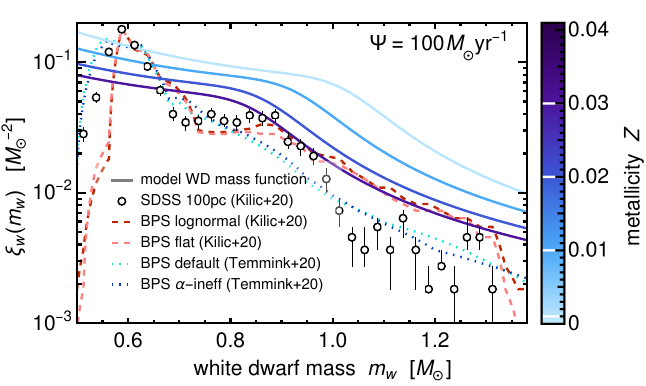}
    \includegraphics[width=\linewidth]{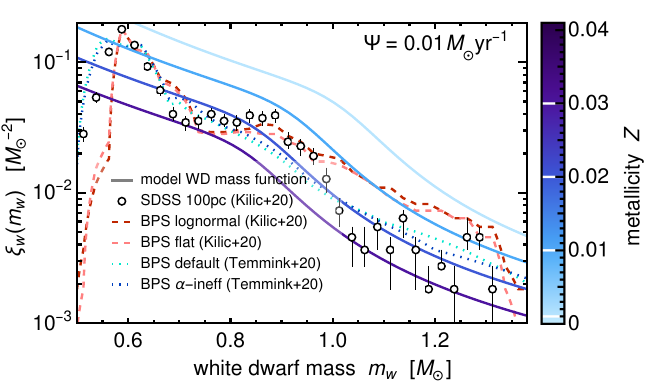}
    \caption{Same as Fig.~\ref{fig:xiwmw}, but for top-heavy (top panel) and top-light (bottom panel) stellar IMFs, corresponding to the SFRs in supermassive galaxies at star formation noon and low-mass galaxies, respectively.}
    \label{fig:xiwmw-SFR}
\end{figure}

In Fig.~\ref{fig:xiwmw-SFR}, we show the WD mass function when SFR is extremely high ($\Psi=100M_{\odot}$yr$^{-1}$, top panel) and extremely low ($\Psi=0.01M_{\odot}$yr$^{-1}$, bottom panel) and the top-part of the stellar IMF is SFR-modulated according to %Fontanot et al. (2017)
Ref.~\cite{2017MNRAS.464.3812F} (see Fig.~\ref{fig:IGIMF}). These two extreme cases correspond to average SFRs of ultramassive ($M_{\star}\sim 10^{11.5}M_{\odot}$) galaxies at star-formation noon and typical low-mass ($M_{\star}\sim 10^8M_{\odot}$) galaxies, respectively, as can be checked comparing Fig.~\ref{fig:SFH}. 
It should be remembered that ultramassive galaxies are metal-rich ($Z\gtrsim 0.03$), while low-mass galaxies are metal-poor ($Z\lesssim 0.01$). Thus, the typical WD mass function of ultramassive galaxies would be the dark blue curve in Fig.~\ref{fig:xiwmw-SFR} top panel, and that of low-mass galaxies either of the lighter blue curves labeled $Z=0.001$ or $Z=0.01$ in Fig.~\ref{fig:xiwmw-SFR} bottom panel. Comparing then the expected WD mass functions for supermassive and low-mass galaxies, it can be noticed that both systems produce approximately the same number of ultramassive ($m_w\gtrsim 1.3M_{\odot}$) WDs per solar mass of stars formed, while low-mass galaxies produce roughly a factor of $\sim 2$ more type Ia supernova progenitors ($m_w\sim 1M_{\odot}$) and a factor of $\sim 4$ more low-mass WDs per solar mass of stars formed than ultramassive galaxies.
This behaviour is explained by the combined effects of IMF and IFMR. For low-mass galaxies, low-metallicity implies bottom-light IMF and high-normalization IFRM, both leading to more WDs per solar mass of stars formed, while low-SFR implies top-light IMF leading to less WDs per solar mass of stars formed and a more steeper mass function slope, which explains why relatively more low-mass WDs are formed than in ultramassive galaxies.

%%%%%%%%%%%%%%%%%%%%%%%%%%%%%%%%%%%%%%%%%%%%%%%%%%%%%%%%%
%%%%%%%%%%%%%%%%%%%%%%%%%%%%%%%%%%%%%%%%%%%%%%%%%%%%%%%%%
\bibliography{biblio}% Produces the bibliography via BibTeX.

\end{document}